\documentclass[twocolumn]{aastex63}
\newcommand{\lya}{Ly$\alpha$}
\newcommand{\oiii}{[O\,{\sc iii}]}

\newcommand{\hb}{H$\beta$}

\newcommand{\w}{$W_r$~}
\newcommand{\kms}{km s$^{-1}$}
\newcommand{\z}{$z$}

\accepted{2024 January 28}

\shorttitle{ASPIRE Abs Galaxy EoR}
\shortauthors{Zou et al.}

\begin{document}

\title{A SPectroscopic survey of biased halos In the Reionization Era (ASPIRE): Impact of Galaxies on the CGM Metal Enrichment at $z >$ 6 Using the JWST and VLT}

\correspondingauthor{Siwei Zou}
\email{siwei1905@gmail.com}

\author[0000-0002-3983-6484]{Siwei Zou}
\affiliation{Department of Astronomy, Tsinghua University, Beijing 100084, China}

\author[0000-0001-8467-6478]{Zheng Cai}
\affiliation{Department of Astronomy, Tsinghua University, Beijing 100084, China}

\author[0000-0002-7633-431X]{Feige Wang}
\affiliation{Steward Observatory, University of Arizona, 933 N Cherry Avenue, Tucson, AZ 85721, USA}

\author[0000-0003-3310-0131]{Xiaohui Fan}
\affiliation{Steward Observatory, University of Arizona, 933 N Cherry Avenue, Tucson, AZ 85721, USA}

\author[0000-0002-6184-9097]{Jaclyn B. Champagne}
\affiliation{Steward Observatory, University of Arizona, 933 N Cherry Avenue, Tucson, AZ 85721, USA}

\author[0000-0002-7054-4332]{Joseph F. Hennawi}
\affiliation{Department of Physics, University of California, Santa Barbara, CA 93106-9530, USA}
\affiliation{Leiden Observatory, Leiden University, Niels Bohrweg 2, NL-2333 CA Leiden, Netherlands}

\author[0000-0002-4544-8242]{Jan-Torge Schindler}
\affiliation{Hamburger Sternwarte, Universität Hamburg, Gojenbergsweg 112, D-21029 Hamburg, Germany}

\author[0000-0002-6822-2254]{Emanuele Paolo Farina}
\affiliation{Gemini Observatory, NSF’s NOIRLab, 670 N A’ohoku Place, Hilo, HI 96720, USA}

\author[0000-0001-5287-4242]{Jinyi Yang}
\altaffiliation{Strittmatter Fellow}
\affiliation{Steward Observatory, University of Arizona, 933 N Cherry Avenue, Tucson, AZ 85721, USA}

\author[0000-0001-9840-4959]{Kohei Inayoshi}
\affiliation{Department of Astronomy, School of Physics, Peking University, Beijing 100871, China}
\affiliation{Kavli Institute for Astronomy and Astrophysics, Peking University, Beijing 100871, China}

\author[0000-0002-2931-7824]{Eduardo Ba\~nados}
\affiliation{Max-Planck-Institut f\"{u}r Astronomie, K\"{o}nigstuhl 17, D-69117 Heidelberg, Germany}

\author[0000-0001-8582-7012]{Sarah E.I. Bosman}
\affiliation{Institute for Theoretical Physics, Heidelberg University, Philosophenweg 12, D–69120, Heidelberg, Germany}
\affiliation{Max-Planck-Institut f\"{u}r Astronomie, K\"{o}nigstuhl 17, D-69117 Heidelberg, Germany}

\author[0000-0001-5951-459X]{Zihao Li}
\affiliation{Department of Astronomy, Tsinghua University, Beijing 100084, China}

\author[0000-0001-6052-4234]{Xiaojing Lin}
\affiliation{Department of Astronomy, Tsinghua University, Beijing 100084, China}

\author[0000-0003-0111-8249]{Yunjing Wu}
\affiliation{Department of Astronomy, Tsinghua University, Beijing 100084, China}

\author[0000-0002-4622-6617]{Fengwu Sun}
\affiliation{Steward Observatory, University of Arizona, 933 N Cherry Avenue, Tucson, AZ 85721, USA}

\author[0000-0002-7532-1496]{Ziyi Guo}
\affiliation{School of Astronomy and Space Science, Nanjing University, Nanjing, 210023, China}
\affiliation{Key Laboratory of Modern Astronomy and Astrophysics (Nanjing University), Ministry of Education, Nanjing 210023, China}

\author[0000-0001-5829-4716]{Girish Kulkuarni}
\affiliation{Tata Institute of Fundamental Research, Homi Bhabha Road, Mumbai 400005, India}


\author[0000-0003-4750-0187]{M\'elanie Habouzit}
\affiliation{Max-Planck-Institut f\"{u}r Astronomie, K\"{o}nigstuhl 17, D-69117 Heidelberg, Germany}
\affiliation{Zentrum f\"ur Astronomie der Universität Heidelberg, ITA, Albert-Ueberle-Str. 2, D-69120 Heidelberg, Germany}

\author[0000-0003-3458-2275]{Stephane Charlot}
\affiliation{Institut d’Astrophysique de Paris, Sorbonne Universit\'e, CNRS, UMR 7095, 98 bis bd Arago, F-75014, Paris, France}

\author[0000-0002-7636-0534]{Jacopo Chevallard}
\affiliation{Scientific Support Office, Directorate of Science and Robotic Exploration, ESA/ESTEC, Keplerlaan 1, NL-2201 AZ Noordwijk, the Netherlands}

\author[0000-0002-7898-7664]{Thomas Connor}
\affiliation{Center for Astrophysics $\vert$\ Harvard\ \&\ Smithsonian, 60 Garden St., Cambridge, MA 02138, USA}

\author[0000-0003-2895-6218]{Anna-Christina Eilers}
\affiliation{MIT Kavli Institute for Astrophysics and Space Research, 77 Massachusetts Ave., Cambridge, MA 02139, USA}

\author[0000-0003-4176-6486]{Linhua Jiang}
\affiliation{Department of Astronomy, School of Physics, Peking University, Beijing 100871, China}
\affiliation{Kavli Institute for Astronomy and Astrophysics, Peking University, Beijing 100871, China}

\author[0000-0002-5768-738X]{Xiangyu Jin}
\affiliation{Steward Observatory, University of Arizona, 933 N Cherry Avenue, Tucson, AZ 85721, USA}

\author[0000-0001-6874-1321]{Koki Kakiichi}
\affiliation{Department of Physics, University of California, Santa Barbara, CA 93106-9530, USA}

\author[0000-0001-6251-649X]{Mingyu Li}
\affiliation{Department of Astronomy, Tsinghua University, Beijing 100084, China}

\author[0000-0001-5492-4522]{Romain A. Meyer}
\affiliation{Max-Planck-Institut f\"{u}r Astronomie, K\"{o}nigstuhl 17, D-69117 Heidelberg, Germany}


\author[0000-0003-4793-7880]{Fabian Walter}
\affiliation{Max-Planck-Institut f\"{u}r Astronomie, K\"{o}nigstuhl 17, D-69117 Heidelberg, Germany}

\author[0000-0002-0123-9246]{Huanian Zhang} 
\affil{Department of Astronomy, Huazhong University of Science and Technology, Wuhan, 430074, China}


\begin{abstract}

We characterize the multiphase circumgalactic medium and galaxy properties at $z$ = 6.0-6.5 in four quasar fields from the James Webb Space Telescope A SPectroscopic survey of biased halos In the Reionization Era (ASPIRE) program. We use the Very Large Telescope/X-shooter spectra of quasar J0305--3150 to identify one new metal absorber at $z$ = 6.2713 with multiple transitions (O~{\sc i}, Mg~{\sc ii}, Fe~{\sc ii} and C~{\sc ii}). They are combined with the published absorbing systems in \citet{davies23a} at the same redshift range to form of a sample of nine metal absorbers at $z$ = 6.03 to 6.49. We identify eight galaxies within 1000 \kms~and 350 kpc around the absorbing gas from the ASPIRE spectroscopic data, with their redshifts secured by \oiii~($\lambda\lambda$4959, 5007) doublets and H$\beta$ emission lines. Our spectral energy distribution fitting indicates that the absorbing galaxies have stellar mass ranging from 10$^{7.2}$ to 10$^{8.8}$~$M_{\odot}$ and metallicity between 0.02 and 0.4 solar. Notably, the $z$ = 6.2713 system in the J0305--3150 field resides in a galaxy overdensity region, which contains two (tentatively) merging galaxies within 350 kpc and seven galaxies within 1 Mpc. We measure the relative abundances of $\alpha$ elements to iron ([$\alpha$/Fe]) and find that the CGM gas in the most overdense region exhibits a lower [$\alpha$/Fe] ratio. Our modeling of the galaxy's chemical abundance favors a top-heavy stellar initial mass function, and hints that we may be witnessing the contribution of the first generation Population \uppercase\expandafter{\romannumeral3} stars to the CGM at the end of reionization epoch.

\end{abstract}


\section{Introduction}

The gaseous halos in the circumgalactic medium (CGM) and the intergalactic medium (IGM) play a critical role in the baryon cycle, subsequently affecting galaxy evolution (for reviews, see \citealt{tum17,per20} and references therein). Exploring the connection between CGM/IGM with the universe's earliest stars, quasars, and galaxies offers insights into the metal enrichment, galaxy assembly, and the origins of supermassive black holes during the reionization epoch (EoR, see reviews by \citealt{fan06,ina20,fan23} and references therein).

The absorbing systems detected in quasar spectroscopy, which arise from intervening gaseous halos towards bright background quasars, provide a sensitive measure of multi-phase gas. Different gas phases are characterized by their density, temperature, and ionization parameters. The multiplicity of metal ions in the absorbing system allows for characterization of the different gas phases such as cold and molecular gas \citep{pet00a,led15,zou18,not18,heintz19}, warm (T $\sim$ 10$^{4}$ K) gas \citep{wolf00a,ste02,raf12,nee19,chen17,zou21,lin23,zhang23,zou24} and warm-hot (T $\sim$ 10$^{5.5-6}$ K) gas \citep{bor14,bur16,davies23a}. In particular, metal enrichment in the IGM/CGM at the end of EoR has been studied by using background quasars at $z >$ 6 \citep{bos18,bana19,coo19,bec19,sim20,zou21,davies23a,davies23b}.

Signatures of the first generation metal-free stars (Population \uppercase\expandafter{\romannumeral3} stars) have been reported in the galactic metal-poor stars (see \citealt{frebel15} for a review), pristine H~{\sc i}-bearing gas such as Damped Lyman-$\alpha$ systems (DLAs) and Lyman-limit-systems (LLSs, \citealt{cooke11,welsh22,zou20,sac23,chris23}) and high redshift galaxies (e.g., \citealt{maio23}). Detections of pristine gas around high-redshift galaxies are promising for our understanding of the formation history of the first objects. Previous detections of gaseous halos and their associated galaxy environments have been largely limited by the capabilities of ground-based telescopes, especially at $z>$ 6.

JWST opens a new era for discovering galaxies---particularly dwarf galaxies at high redshift (e.g., \citealt{atek22,curtis23,end22,mat23,fengwu23,jades_a,fuji23}). Combining the data from distant galaxies detected by JWST with the absorbing gas in their CGM can enhance our understanding of chemical enrichment in the CGM/IGM and its connection to the formation of the first objects.

The structure of this paper is presented as follows: In Section \ref{sec:survey_reduction}, we introduce the A SPectroscopic survey of biased halos In the Reionization Era (ASPIRE) program. Section \ref{sec:absorber_galaxy} details the absorbers and the detection of surrounding galaxies. The properties of the absorbing gas and its connection with the galaxies are discussed in Section \ref{sec:discussion}. We use the standard $\Lambda$CDM model in this work and the used cosmological parameters are: $H_0$ = 70 km s$^{-1}$ Mpc$^{-1}$, $\Omega_\Lambda$ = 0.69 and  $\Omega_m$ = 0.31.     


\section{Sample selection and data reduction}\label{sec:survey_reduction}

The ASPIRE survey encompasses 25 quasars in the range (6.5 $< z <$ 6.8), each with high-quality spectra spanning from optical to radio observations (e.g., \citealt{ven19,yang20b,yang21}). These quasar fields are observed using JWST's NIRCam/WFSS. Photometric data are collected in three bandpass filters: F115W, F200W, and F356W. The Grism R spectra are acquired in the F356W filter, encompassing the \oiii~and \hb~emission lines for redshifts approximately between \(5.2\) and \(7.0\). The grism spectra offer a resolving power of \(1000-1500\) at wavelengths of \(3-4 \mu m\). Details about this survey can be found in \citep{wang23,yang23}.

\subsection{JWST/NIRCam data reduction}
We perform data reduction for the JWST/NIRCam and WFSS data using version 1.10.2 of the JWST Calibration Pipeline (CALWEBB). The official CALWEBB pipeline processes data products in three stages. \citet{wang23} provides detailed data reduction steps, and we offer a brief overview here. In the NIRCAM direct image processing, the 1/f noise pattern is first eliminated. For data calibration, we employ reference files (jwst 1080.pmap) from version 11.16.21 of the standard Calibration Reference Data System (CRDS). All imaging data in this study align with the Gaia data release 3. For the WFSS data, the CALWEBB Stage 1 pipeline calibrates detector-level signals and performs ramp fitting for individual NIRCam WFSS exposures. To mitigate the 1/f noise, we subtract the calibrated data spectra column-wise as the spectra disperse row-wise for Grism-R. It's worth noting that we achieve background subtraction by creating background models based on all ASPIRE observations taken at similar times; we then scale and subtract these models from individual WFSS exposures.

A critical step in the data reduction process for slitless spectra involves the tracing model. We adopt the method outlined in \citet{fengwu22}. This model traces point sources observed in the Large Magellanic Cloud field. The sensitivity functions, as determined from both the ERO calibration and the Cycle-1 calibration, align with an accuracy surpassing 98\% across most of the wavelength range. Ultimately, we extract both the 2D and 1D spectra from the catalog of detected imaging. The depth reaches a 5$\sigma$ line luminosity limit of 9.9 $\times$ 10$^{41}$ erg s$^{-1}$ at $z$ = 6.6.

\subsection{VLT/X-shooter data reduction}

In this work, we analyze the connection between absorbing gas and galaxies using four quasar sightlines (J0305--3150, J0226+0302, J0224--4711, and J0923+0402) for which we have both VLT/X-shooter and ASPIRE JWST/WFSS data. The latter three sightlines are part of the XQR-30 sample \citep{dod23}. The absorber catalog can be found in \citet{davies23a}. The quasar J0305--3150 is included in the X-shooter/ALMA Sample of Quasars in the Epoch of Reionization \citep{jt20,farina22}. We reduce the X-shooter near-infrared (NIR) spectra using the Python Spectroscopic Data Reduction Pipeline, PypeIt \citep{pypeit}, with the reduction details presented in \citet{jt20}.

\section{Data analysis}\label{sec:absorber_galaxy}

\subsection{Detection of absorbers and associated \oiii~emitters }
We identify metal absorbers from the VLT/X-shooter spectra. Due to the sample size, we manually normalize the reduced X-shooter spectra using a spline function and subsequently conduct a visual inspection of the absorbers. We measure the column densities and Doppler parameters by fitting the absorption lines with a Voigt profile, utilizing the VPFIT code \citep{kro18}. We added a 10\% uncertainty from the continuum fitting when calculating the column densities of different ions. We also incorporate absorbing systems at $z > 6.0$ along three XQR-30 sightlines (J0226+0302, J0224--4711 and J0923+0402) as detailed in \citet{davies23a}. Specifically, we measure the upper limits of Fe~{\sc ii} column density, $N_{\text{Fe~{\sc ii}}}$, by varying the Doppler parameter $b$ to obtain a reasonable fitting curve.

\begin{figure*}[htb!]

\gridline{
\fig{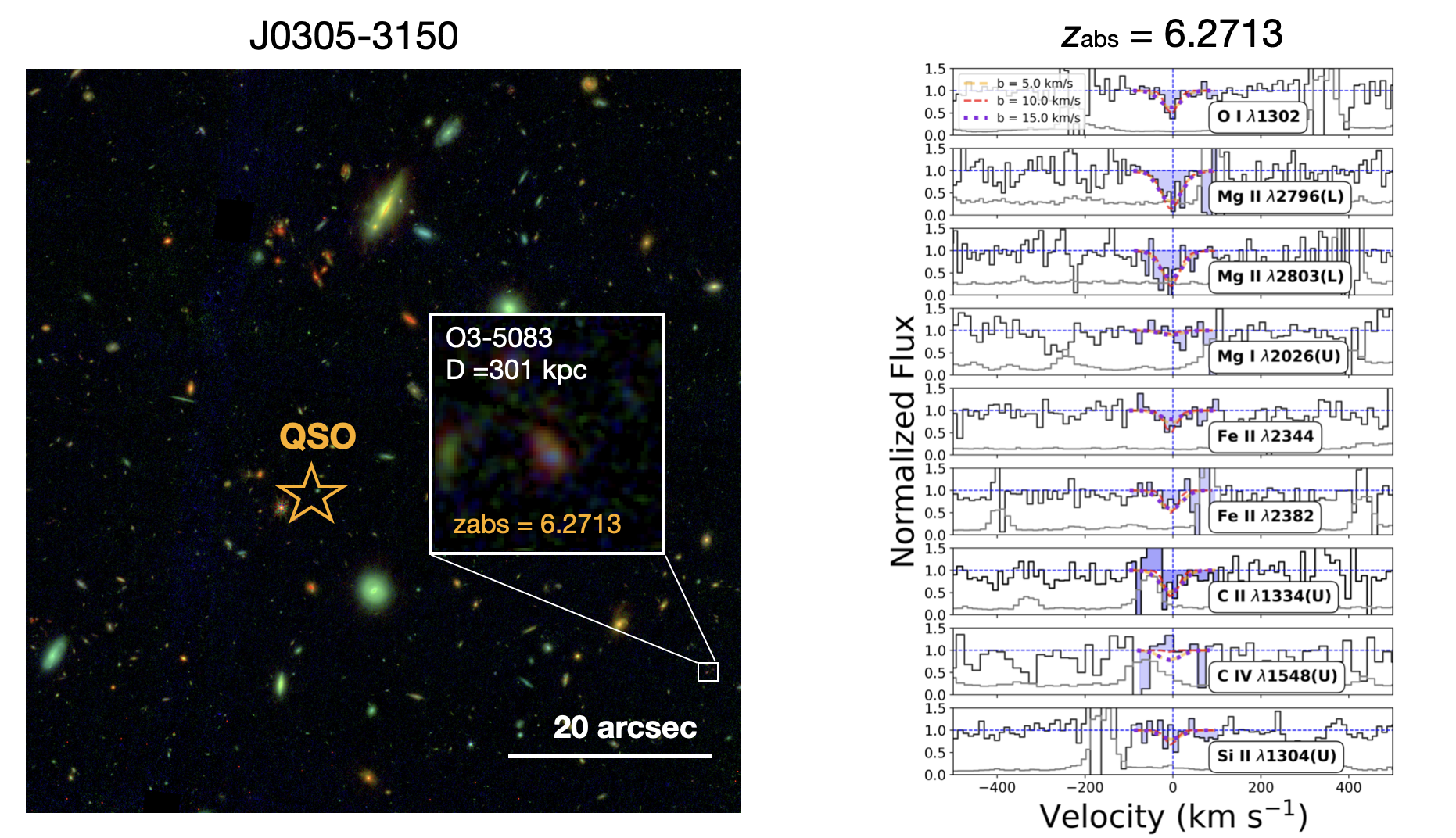}{1.0\textwidth}{}
}

\caption{The absorbers detected at $z$ = 6.2713 towards quasar J0305--3150 and associated \oiii~emitter candidate. The purple, orange, and red curves represent the fitting of absorption lines in this system with a Voigt profile and Doppler parameters of 5, 10, and 15 \kms, respectively. The letters U and L represent the upper and lower limits on the column density measurement of the lines. The galaxy candidates detected within 350 kpc and 1000 km s$^{-1}$ of the absorbing gas are presented in the zoom-in RGB imaging boxes.}\label{fig:j0305-field}
\end{figure*}

\begin{figure*}[htb!]

\gridline{
\fig{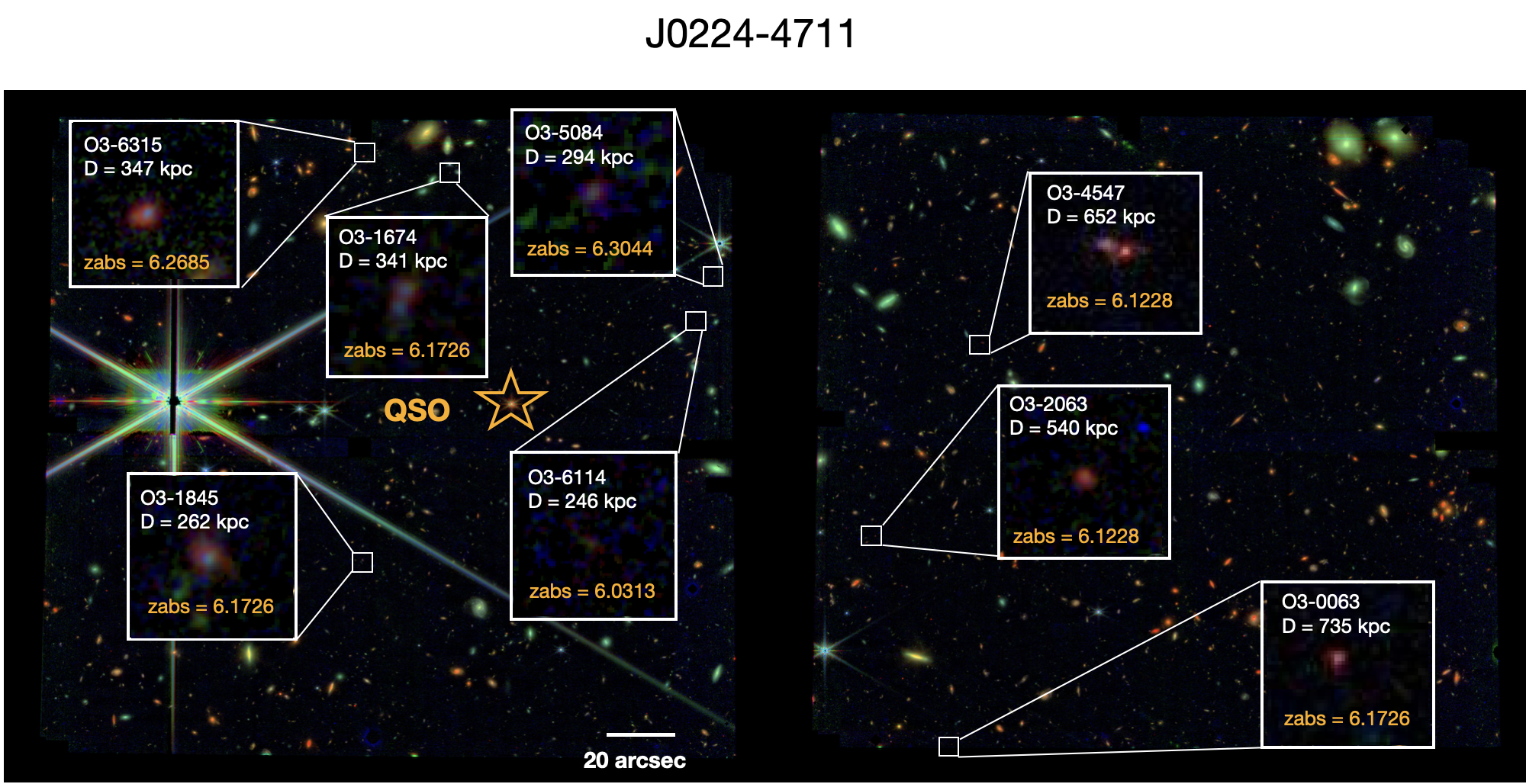}{1.0\textwidth}{}
}

\gridline{
\fig{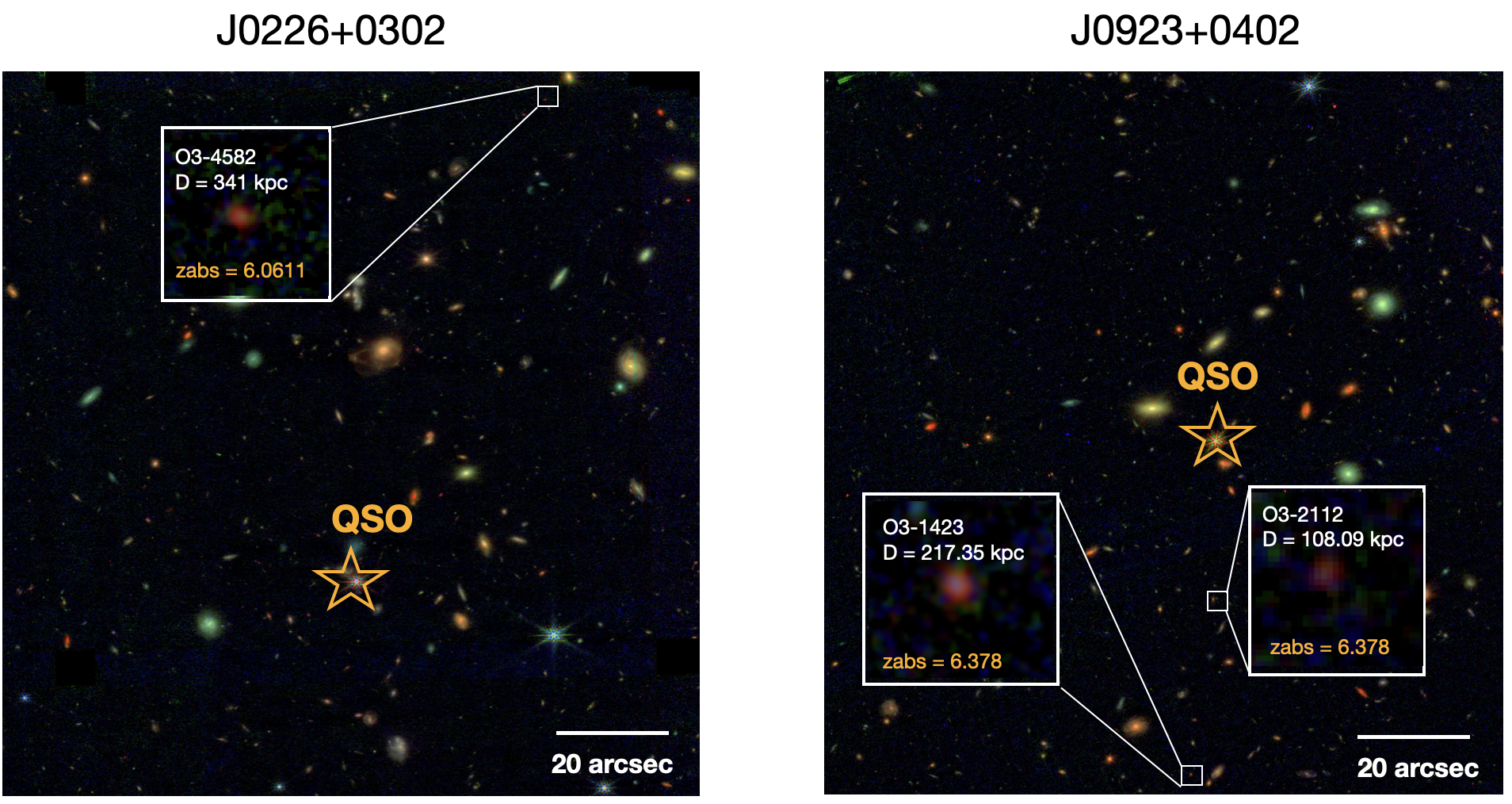}{1.0\textwidth}{}
}

\caption{We include the absorbing systems at $z>$ 6.0 in \citet{davies23a} along the quasar sightlines J0226+0302, J0923+0402 and J0224--4711. The galaxies detected within 350 kpc and 1000 km s$^{-1}$ of the absorbing gas are presented in the zoom-in RGB imaging boxes.}\label{fig:other-field}
\end{figure*}

We then search for the \oiii~emitters within a velocity window of $\pm$1000 \kms~and an impact parameter of 350 kpc from the absorbing gas. To determine the expected number of \oiii~emitters within this volume, we refer to the \oiii~luminosity function at  $z \sim 6.2$ from \citet{fengwu22}. Considering our detection limit (with an \oiii~luminosity threshold of 10$^{42.3}$  erg s$^{-1}$), the count is 8$\times10^{-5}$. The detection of one galaxy or more in this context suggests its potential association with the absorbing gas. We use the algorithm described in \citet{wang23} to automatically search for the \oiii~emitters having a peak S/N $>$ 3.0. We then visually checked all the candidates and confirmed the final \oiii~targets. We present the detected absorber and associated galaxy candidates in Figures \ref{fig:j0305-field}, \ref{fig:other-field} and Appendix Figures \ref{fig:vpfit-xqr30-a1}-\ref{fig:vpfit-xqr30-a4}. The full 2D and 1D spectra of all the \oiii~emitters can be found in the Appendix Figure \ref{fig:o3_spec}. Characteristics of the detected absorbers and galaxy candidates are presented in Table \ref{table:absorber-galaxy}.

\vspace{0.1cm}
\noindent{\bf J0305--3150}\\
\noindent We detect one absorbing system having O~{\sc i}, C~{\sc ii}, Mg~{\sc ii}, and Fe~{\sc ii} at $z$ = 6.2713. The rest-frame equivalent widths ($W_r$) of O~{\sc i}($\lambda$1302), Mg~{\sc ii}($\lambda$2796), Mg~{\sc ii}($\lambda$2803) and Fe~{\sc ii}($\lambda$2344) lines are 0.069$\pm$0.022 \AA, 0.32$\pm$0.12 \AA, 0.30$\pm$0.14 \AA~and 0.17$\pm$0.10 \AA, respectively. The $W_r$ ratio of the Mg~{\sc ii} doublet $W_r^{\lambda2803}/W_r^{\lambda2796}$ is 0.94, indicating strong saturation in the Mg~{\sc ii} absorption. With a Doppler parameter of $b = 10-15$ \kms, the log $N_{\text{O~{\sc i}}}$ ranges from 14.26 to 14.38, falling within the slightly saturated region on the curve of growth(COG). The Fe~{\sc ii} is slightly saturated, given that log $N_{\text{Fe~{\sc ii}}}>$ 13.2. There is no significant ($W_r > 0.1$ \AA) detection of Si~{\sc ii} and C~{\sc iv} lines. The Si~{\sc iv} lines are significantly contaminated by sky lines. The C~{\sc ii} ($\lambda$1334) is affected by the sky lines, so we measure its column density upper limit in Table \ref{table:absorber-galaxy}. We plot the metal lines and fitting curve with a Voigt profile in Figure \ref{fig:j0305-field}, the right panel.


We detect one possible \oiii~emitter pair at impact parameter $D\sim$ 298 kpc (proper distance) and velocity offset $\Delta v$ = 708 \kms: ASPIRE-J0305M31-O3-5083 (see Figure \ref{fig:j0305-field}). This emitter displays a merging feature in both the 2D and 1D spectra: two sources are identified using SEXtractor. Notably, seven other \oiii~emitters reside within $D$ = 1 Mpc. The ASPIRE-J0305M31-O3-0623 and ASPIRE-J0305M31-O3-0905 also exhibit a merging feature. \citet{wang23} present two galaxy overdensity fields at $z$ = 5.2 and 6.2. This metal-enriched absorbing gas resides in the overdense region at $z\sim$ 6.2 (a discussion of the $z$ = 5.2 absorbing system is presented in \citealt{wu23}). 

\vspace{0.1cm}
\noindent{\bf XQR-30 absorbers}
\vspace{0.1cm}

\noindent{\bf J0226+0302}\\
\noindent We use the absorbing system at $z = $ 6.06111 detected in the QSO slightline J0226+0302 (target name PJ036+03 in XQR-30 sample) to search for galaxy counterpart candidates. This absorbing system contains O~{\sc i}, Mg~{\sc ii} and Si~{\sc ii}. The $W_r$ ratio between the Mg~{\sc ii} doublet $W_r^{\lambda2803}/W_r^{\lambda2796}$ is 0.64$\pm$0.05, indicating the Mg~{\sc ii} is slightly saturated. The O~{\sc i}, C~{\sc ii}, C~{\sc iv} and Si~{\sc ii} lines are not saturated. No Fe~{\sc ii} line are detected. We re-measure the $N_{\text{Fe~{\sc ii}}}$ and obtain its upper limit in Table \ref{table:absorber-galaxy}. From our WFSS data, one \oiii~emitter: ASPIRE-J0224M47-O3-4582, is detected at $D = $ 341 kpc and $\Delta v$ = 155 \kms. 



\vspace{0.1cm}
\noindent{\bf J0224--4711}\\
\noindent We use the absorbing systems at $z\geq$ 6.0 (6.03133, 6.12279, 6.17255, 6.26848, 6.30443, and 6.4820) from \citet{davies23a} to search for galaxy counterpart candidates. The O~{\sc i} is detected in the system at $z = 6.12279$. We refit the metal lines with a Voigt profile and vary the $b$ value to obtain a reasonable profile. The results are presented in Figure \ref{fig:vpfit-xqr30-a2}. With $b = 10-15.8$ \kms, the log $N_{\text{Fe~{\sc ii}}}$ varies within 0.04, and log $N_{\text{O~{\sc i}}}$ varies between 14.67$\pm$0.11 to 14.46$\pm$0.06. The Fe lines are likely not saturated, and the O~{\sc i} is tentatively partially saturated. The $W_r^{\lambda2803}/W_r^{\lambda2796}$ is 0.67$\pm$0.02, indicating partial saturation in the Mg~{\sc ii} doublet. The C~{\sc iv} is contaminated by sky lines, so we measure an upper limit for the $N_{\text{C~{\sc iv}}}$. The C~{\sc ii}($\lambda$1334) line is slightly saturated. We find no galaxies within 350 kpc for this system. There is one \oiii~emitter pair, ASPIRE-J0224M47-O3-4547, at $D\sim650$ kpc, and one isolated \oiii~emitter, ASPIRE-J0224M47-O3-2063, at $D = 540$ kpc.

For the system at $z = 6.26848$, the $W_r^{\lambda2803}/W_r^{\lambda2796}$ is $0.67 \pm 0.03$, indicating partial saturation in the Mg~{\sc ii} doublet. The O~{\sc i} is not reported as detected \citep{davies23a} given its strong blending with the Mg~{\sc ii} absorption at $z = 2.3$. There are no Fe~{\sc ii} lines detected, so we measured the upper limit of $N_{\text{Fe~{\sc ii}}}$. The C~{\sc ii} is contaminated by skylines. The second component of the C~{\sc ii} line at $\Delta v \sim 45$ km s$^{-1}$ likely blended with another line, so we only fit the first component. The C~{\sc iv} is partially contaminated by the sky lines. One galaxy ASPIRE-J0224M47-O3-6114 is detected at $D = $ 246 kpc and $\Delta v =$ 41.5 km s$^{-1}$. One galaxy ASPIRE-J0224M47-O3-6315 is detected at $D$ = 347 kpc and $\Delta v$ = 1047 \kms. 

For the system at $z = 6.03133$, only C~{\sc iv} is reported as detection in \citep{davies23a}. The C~{\sc ii}($\lambda$1334) line also presents but likely blended with other lines. We thus give an upper limit on the $N_{\text{C~{\sc ii}}}$. We re-measured the upper limit for $N_{\text{Si~{\sc iv}}}$ and $N_{\text{Fe~{\sc ii}}}$. One galaxy ASPIRE-J0224M47-O3-6114 is detected at $D = $ 246 kpc and $\Delta v =$ 41.5 km s$^{-1}$.

For the system at $z = 6.17255$, only C~{\sc iv} is detected and no detection for C~{\sc ii} and Fe~{\sc ii} lines, so we measure the $N_{\text{C~{\sc ii}}}$, $N_{\text{Fe~{\sc ii}}}$ upper limits for this system. One galaxy ASPIRE-J0224M47-O3-1674 is detected at $D = $ 341 kpc and $\Delta v =$ 320.7 km s$^{-1}$.

For the system at $z = 6.30443$, Mg~{\sc ii}, C~{\sc ii}, and Si~{\sc ii}($\lambda$1260) lines are reported as detection in \citep{davies23a}. Considering the profile of C~{\sc iv}, we report its $N$ measurement as an upper limit in Table \ref{table:absorber-galaxy}. No Fe~{\sc ii} lines are detected. The $N_{\text{Mg~{\sc ii}}}$ is in the slightly saturated region in the COG, and the ratio of $W_r^{\lambda2803}/W_r^{\lambda2796}$ is 0.63$\pm$ 0.02. One galaxy ASPIRE-J0224M47-O3-5084 is detected at $D = $ 194 kpc and $\Delta v =$ 126.0 km s$^{-1}$.

For the $z =$ 6.4820 system having only C~{\sc iv} absorption, we do not find any galaxy candidates within 1 Mpc.

\vspace{0.1cm}
\noindent{\bf J0923+0402}\\
\noindent This target is in the XQR-30 sample. We select the absorbing systems at $z$ = 6.378 from \citet{davies23a} to search for galaxy counterparts. We detect two \oiii+\hb~emitters (ASPIRE-J0923P04-O3-2112 and 1423) at $D$ = 108 and 217 kpc, respectively. This absorbing system has O~{\sc i}, Mg~{\sc ii}, C~{\sc ii} and Si~{\sc ii} absorption. The O~{\sc i}($\lambda$1302) equivalent width is 0.074$\pm$0.004 \AA~and the $N_{\text{O~{\sc i}}}$ is in the saturated region in the COG. The $W_r$ ratio between the Mg~{\sc ii} doublet $W_r^{\lambda2803}/W_r^{\lambda2796}$ is 0.77$\pm$0.05, indicating a partial saturation. No Fe~{\sc ii} lines are detected and the C~{\sc iv} lines are significantly contaminated by the skylines.  

\subsection{SED fitting of [O\sc{iii}]+H$\beta$ emitters}
We use the tool \textsc{BEAGLE} (BayEsian Analysis of GaLaxy sEds; \citealt{beagle}) to perform spectral energy distribution (SED) fitting for the absorbing galaxies. \textsc{BEAGLE} determines the most probable properties of a galaxy---including the ionization parameter ($U$), metallicity, SFR, $M_*$, age, and UV slope $\beta$---based on its photometric and spectroscopic data. The initial mass function (IMF), which describes the distribution of stellar masses (e.g., \citealt{sal95,kroupa01,cha03}), is fixed using the Galactic-disc IMF from \citet{cha03}. Specifically, the iron abundance [Fe/H] ranges from [--1, 0.2], and the $\alpha$ elements-to-iron abundance ratio [$\alpha$/Fe] spans [0, 0.4]. The photoionization models for star-forming galaxies are presented in \citet{gut16}. 

For the SED fitting, we incorporate the observed photometry in F115W, F200W, and F356W bands, as well as the spectroscopically observed fluxes of \oiii~and \hb~lines. For J0305--3150, we alaso include archived HST photometry from the F105W, F125W, F160W bands (PropID 15064, PI: Casey). We adopt a `constant' star formation history and use the Small Magellanic Cloud (SMC) UV extinction law. Our priors on galaxy stellar mass are set as $5.0 \leq \log M_* / M_\odot \leq 12$, on metallicity as $-4.0 \leq \log Z / Z_\odot \leq 1.0$, and on the ionization parameter as $-4.0 \leq \log U_\odot \leq -1.0$. The results suggest stellar masses for our galaxies in the range of $7.0 \leq \log M_* / M_\odot \leq 9.0$ (with a median value of log $M_*/M_{\odot}$ = 8.1) and metallicity in the range of $-1.0 \leq \log Z / Z_\odot \leq 0.0$ (as illustrated in Appendix Figure \ref{fig:sed}). It is worth noting that in \textsc{BEAGLE}, the IMF remains fixed, ensuring that the contributions from different stellar masses and generations are constant.

To quantify the uncertainties in the SED fitting, we randomly selected 100 galaxies from the JAGUAR mock galaxy catalog \citep{jaguar}\footnote{https://fenrir.as.arizona.edu/jwstmock/index.html} and performed SED modeling on these mock galaxies using the same parameter settings as mentioned above. The \oiii~5007 line flux, redshifts, and stellar mass of these galaxies are consistent with those in our sample. We used their F115W, F200W, and F356W flux densities, assuming a similar photometric uncertainty to our detection depth. Different star formation histories (SFH), `constant,' and `ssp,' are adopted in the tests. We compare the best-fit parameters with the intrinsic values in the mock catalog. We find that if a `constant' SFH is adopted, the standard deviation between the fitted and intrinsic stellar properties---stellar mass, age, metallicity, SFR and $M_{UV}$---is 0.39 dex, 234 Myr, 0.30 dex, 9.5 $M_{\odot}$/yr and 0.17 dex, respectively. This offset in stellar mass, age, metallicity and SFR are 0.56 dex, 173 Myr, 0.25 dex, and 11.2 $M_{\odot}$/yr, respectively, by adopting an `ssp' SFH. We detail the results with added uncertainty on the measurements of the SED fitting for absorbing galaxies in the Table \ref{table:absorber-galaxy} and Appendix Figure \ref{fig:sed}. 


\begin{table*}
\centering
  \caption{Measurements of the properties of absorbers and associated galaxy candidates in this study are provided. In the top table, we re-measure the upper limits of ion column density using the same $b$ value as in \citet{davies23a}; these values are highlighted in bold font. In the lower table, the stellar mass, star formation rate, metallicity, ionization parameter, and age are results from the SED fitting using BEAGLE. \label{table:absorber-galaxy} }
\begin{tabular}{ccccccccccc}
\hline
 &&&&&{\bf Absorbers}&&&&& \\
\hline
Fields & $z_{abs}$   & log $N_{\text{O~{\sc i}}}$  & log $N_{\text{Mg~{\sc ii}}}$ & log $N_{\text{Fe~{\sc ii}}}$& log $N_{\text{C~{\sc ii}}}$ & log $N_{\text{C~{\sc iv}}}$ & log $N_{\text{Si~{\sc ii}}}$ &[$\alpha$/Fe] & $b$  \\
 & &cm$^{-2}$ & cm$^{-2}$ & cm$^{-2}$& cm$^{-2}$& cm$^{-2}$& cm$^{-2}$&  & km s$^{-1}$\\
 \hline
J0305M31 & 6.2713  &  14.37$\pm$0.25   & $>$14.24  & 13.47$\pm$0.25 &  $<$14.21  & $<$13.17 & $<$ 13.64 & --0.29$\pm$0.34 & 10.0  \\
J0226P03\footnote{\label{xqr30} Quasars are included in the XQR-30 sample} & 6.0611  & 13.89$\pm$0.05  & 12.80$\pm$0.14 & {\bf $<$12.64}\footnote{\label{davb} Re-measured column density with the same $b$ value in \citet{davies23a}.}  & 13.48$\pm$0.09  & {\bf 12.89$\pm$0.18}\textsuperscript{\ref{davb}} & 13.04$\pm$0.12 & $>$0.06 & 8.5 \\
J0923P04\textsuperscript{\ref{xqr30}} & 6.378   & $>$14.68   &  $>$12.88   & {\bf $<$ 12.80}\textsuperscript{\ref{davb}} &  13.79$\pm$0.32  & \nodata  & 13.06$\pm$0.07 & $>$0.69 & 5.9 \\
J0224M47\textsuperscript{\ref{xqr30}}  & 6.03133 & \nodata        & \nodata   & {\bf $<$12.58}\textsuperscript{\ref{davb}} &  {\bf $<$ 13.50}\textsuperscript{\ref{davb}}  & 13.71$\pm$0.07  & \nodata  & $>$0.41 & 13.9 \\
J0224M47\textsuperscript{\ref{xqr30}} & 6.12279 & $>$14.45  & $>$13.15  & 12.83$\pm$0.06 &  $>$13.97  & $<$13.15   &  \nodata  &  $>$0.43  & 15.8\\
J0224M47\textsuperscript{\ref{xqr30}} & 6.17255 & \nodata         & {\bf $<$ 12.00}\textsuperscript{\ref{davb}}   & {\bf $<$ 12.31}\textsuperscript{\ref{davb}} & {\bf $<$ 13.09}\textsuperscript{\ref{davb}}  & 13.99$\pm$0.04 & \nodata  &  $>$0.76 & 34.4\\
J0224M47\textsuperscript{\ref{xqr30}} &  6.26848 & \nodata         & $>$13.08  & $<$ 12.33  &  13.61$\pm$0.11 & {\bf $<$13.92}\textsuperscript{\ref{davb}}  & \nodata  & $>$0.35 & 13.7\\
J0224M47\textsuperscript{\ref{xqr30}} & 6.30443 & \nodata         & 12.55$\pm$0.11 & {\bf $<$ 12.38}\textsuperscript{\ref{davb}} & 13.67$\pm$0.21 & {\bf $<$13.69}\textsuperscript{\ref{davb}}   & 12.11$\pm$0.05 &  $>$0.70 & 6.4\\
J0224M47\textsuperscript{\ref{xqr30}} & 6.4821 & \nodata          & \nodata & \nodata & \nodata &  13.20$\pm$0.07   & \nodata & \nodata & 87.0\\
 
\hline
\end{tabular}

\vspace{0.7cm}
\setlength\tabcolsep{2pt}
\begin{tabular}{lcccccllllll}
\hline
 &&&&&{\bf Galaxies}&&&&& \\
\hline
\small

Galaxy & $z_{g}$  & RA &  DEC & $D$ & $\Delta$ v   & log $M_*$/$M_\odot$ & SFR  & log Z/Z$_\odot$ & log $U$ & Age & $M_{UV}$ \\
 &  & &   & (kpc) & (\kms)  & &  ($M_{\odot}$/yr) &  & &  (Myr) &     \\

J0305M31-O3-5083 & 6.259 & 46.30453 & --31.85411 & 301 & 708.3 & 8.09$^{+0.50}_{-0.50}$ & 11.6$^{+11.03}_{-11.33}$ & --1.00$^{+1.05}_{-0.48}$ & --2.08$^{+1.02}_{-1.05}$  & 16$^{+280}_{-243}$ & --20.43$^{+0.22}_{-0.22}$\\ 
J0226P03-O3-4582 & 6.059 & 36.49517 & 3.03944 & 341 & 155.1  & 7.67$^{+0.53}_{-0.47}$ & 4.64$^{+11.22}_{-10.29}$ & --1.09$^{+0.72}_{-0.53}$ & --2.07$^{+1.20}_{-0.76}$  & 2$^{+236}_{-235}$ & --19.52$^{+0.35}_{-0.39}$ \\ 
J0923P04-O3-2112  & 6.386 & 140.95164 & 4.04869  & 108 &  106.1 &  8.80$^{+0.68}_{-0.88}$ & 0.14$^{+24.89}_{-9.64}$ & --1.58$^{+1.51}_{-0.71}$ & --3.11$^{+1.69}_{-1.19}$ &  31$^{+288}_{-249}$ & --19.83$^{+0.23}_{-0.24}$ \\
J0923P04-O3-1423 & 6.393& 140.95702 & 4.04924 &  217  &  398.0  & 8.08$^{+0.67}_{-0.47}$ & 12.10$^{+16.14}_{-11.45}$ & --1.00$^{+0.78}_{-0.81}$ & --3.04$^{+1.93}_{-0.86}$  & 4$^{+308}_{-235}$ & --20.05$^{+0.19}_{-0.19}$  \\ 
J0224M47-O3-6114  &6.032 & 36.09066 & --47.18788 & 246 & 41.5 & 7.21$^{+0.50}_{-0.48}$ &  1.63$^{+9.91}_{-9.8}$ & --1.32$^{+1.27}_{-0.84}$ & --3.08$^{+1.66}_{-1.01}$  &1$^{+236}_{-235}$ &--17.99$^{+0.25}_{-0.26}$ \\
J0224M47-O3-1845  & 6.152 & 36.12407 & --47.18276  & 262 & 867.2 & 8.71$^{+0.56}_{-0.58}$ & 10.49$^{+18.79}_{-11.89}$ & --0.66$^{+0.87}_{-0.67}$ & --2.40$^{+1.53}_{-1.58}$  &40$^{+427}_{-258}$ & --20.78$^{+0.25}_{-0.25}$  \\  
J0224M47-O3-1674  & 6.167 & 36.09066 & --47.18788 & 341 & 320.7 & 8.24$^{+1.37}_{-0.44}$ &  16.01$^{+11.46}_{-25.46}$ & --0.42$^{+0.48}_{-1.63}$ & --3.94$^{+1.78}_{-0.59}$  & 3$^{+244}_{-235}$ & --19.25$^{+0.26}_{-0.26}$ \\ 
J0224M47-O3-6315  & 6.242 &36.08924 &-- 47.18281  & 347 & 1057.4 & 8.92$^{+0.51}_{-0.50}$ & 16.78$^{+15.31}_{-11.08}$ & --0.72$^{+0.59}_{-0.60}$ & --2.15$^{+0.96}_{-0.82}$  & 24$^{+276}_{-245}$  &--21.32$^{+0.23}_{-0.23}$ \\ 
J0224M47-O3-5084  & 6.308 & 36.09999 & --47.20420  & 294 & 126.0  & 7.39$^{+0.46}_{-0.44}$ &  2.48$^{+9.96}_{-9.78}$ & --1.59$^{+0.73}_{-0.61}$ & --2.60$^{+1.38}_{-1.25}$  &1 $^{+244}_{-244}$ &--19.09$^{+0.25}_{-0.26}$ \\
J0224M47-O3-4547\footnote{\label{d1mpc}impact parameter within 1Mpc}  & 6.130 & 36.10521 & --47.22264  & 652 & 302.0  & 9.21$^{+0.88}_{-0.85}$ &  4.26$^{+11.67}_{-13.71}$ & --1.09$^{+0.45}_{-0.43}$ & --2.36$^{+1.24}_{-1.24}$  &274$^{+481}_{-373}$ &--20.16$^{+0.26}_{-0.27}$ \\ 
J0224M47-O3-2063\textsuperscript{\ref{d1mpc}}  & 6.132 & 36.12200 & --47.21628 & 540  & 277.4  & 7.83$^{+0.48}_{-0.46}$ &  6.78$^{+10.99}_{-10.48}$ & --0.49$^{+0.38}_{-0.42}$ & --1.88$^{+0.78}_{-0.74}$  & 1$^{+235}_{-234}$ & --18.87$^{+0.32}_{-0.31}$ \\  
J0224M47-O3-0063\textsuperscript{\ref{d1mpc}}  & 6.166 & 36.14012 & --47.22079  & 735  & 280.6  & 8.26$^{+0.88}_{-0.79}$ &  6.10$^{+14.64}_{-12.1}$ & --0.92$^{+0.97}_{-0.87}$ & --2.04$^{+1.20}_{-1.38}$  & 78$^{+543}_{-303}$ & --16.57$^{+0.63}_{-0.85}$ \\  

\hline
\end{tabular}
\end{table*}%

\section{Discussion}\label{sec:discussion}

In this section, we discuss the connection between the CGM gas properties (the CGM metal budget and ionization state) and the characteristics of surrounding galaxies at the end of the EoR. We focus on two aspects: {\bf (1)} the metal abundances in the CGM gas and their connection with the star formation history of proximate galaxies; and {\bf (2)} the influence of nearby galaxies as local ionizing sources on the absorbing gas.

\subsection{Relative abundances between iron and $\alpha$-elements in the CGM at $z$ = 6.0 -- 6.5}\label{sec:alpha_fe}



Metals residing in the CGM of a galaxy are delivered out of the galaxy by feedback such as stellar winds, supernovae (SN), and/or AGN feedback \citep{tum17}. Stellar feedback is likely to dominate the baryon cycle in dwarf galaxies during the early cosmic times \citep{cen01}. \citet{cen01} posits that dwarf galaxies formed between $z$ = 7--15 contribute significantly to the universe's metals and energy. Assuming a constant dispersing velocity of $v_{disp}\sim$ 300 km s$^{-1}$ for a gas halo in a galaxy formed at $z = 11.7$ (age $\sim$ 0.36 Gyr), this gas can be pushed out of the host halo ($\sim 10^{10}M_{\odot}$) and travel as far as 1 comoving Mpc within half of the Hubble time (until $z\sim$ 7). Such a dispersion velocity might arise from a sustained starburst or galaxy merger. Therefore, the relative abundances in the CGM gas phase can provide insights into nucleosynthesis and the star formation history of associated dwarf galaxies. Recent findings from the EIGER \citep{eiger_bor} and XQR-30 \citep{davies23b} surveys indeed indicate a rapid metal pollution from galaxies to the IGM at high redshift. 

The $\alpha$-elements are products of Type II supernovae. The Fe-peak elements primarily arise from Type Ia supernovae, which manifest with a delay of about 1 Gyr compared to the $\alpha$-elements \citep{maoz12}. At $z>$ 6, where the Hubble time is less than 1 Gyr, the gas-phase [$\alpha$/Fe] abundance becomes an interesting indicator of the gaseous halo's lifetime. It also offers insights into the star formation history and the process of galaxy assembly.  

For the absorbing gas at $z>$ 6.0 in this work, even though we lack H~{\sc i} column measurements due to H~{\sc i} reionization, we can still determine its relative abundances of $\alpha$ elements over iron using multiple absorption lines. To compute abundances for any two elements (X and Y) based solely on their column densities, we assume
\begin{equation}
[X/Y] = \textrm{log}(N_{\text{X}}/N_{\text{Y}}) - \textrm{log}(N_{\text{X}_\odot}/N_{\text{Y}_\odot}). 
\end{equation}
The solar abundances are taken from \citet{asp09}. For each species, we refer to the dominant ions for its overall column density $N_{\text{X}}$. For example, we have $N_{\text{O}}$ $\sim$ $N_{\text{O~{\sc i}}}$ and $N_{\text{C}}$ $\sim$ $N_{\text{C~{\sc ii}+C~{\sc iv}}}$. We also estimate the $N_{\text{C~{\sc ii}}}$ from $N_{\text{Mg~{\sc ii}}}$ using the empirical relation in \citet{coo19} when C~{\sc ii} is not detected or strongly contaminated: log $N_{\text{C~{\sc ii}}}$ = 0.811$\times$(log $N_{\text{Mg~{\sc ii}}}$--12)+13.09. 

\begin{figure*}[htb!]

\gridline{
\fig{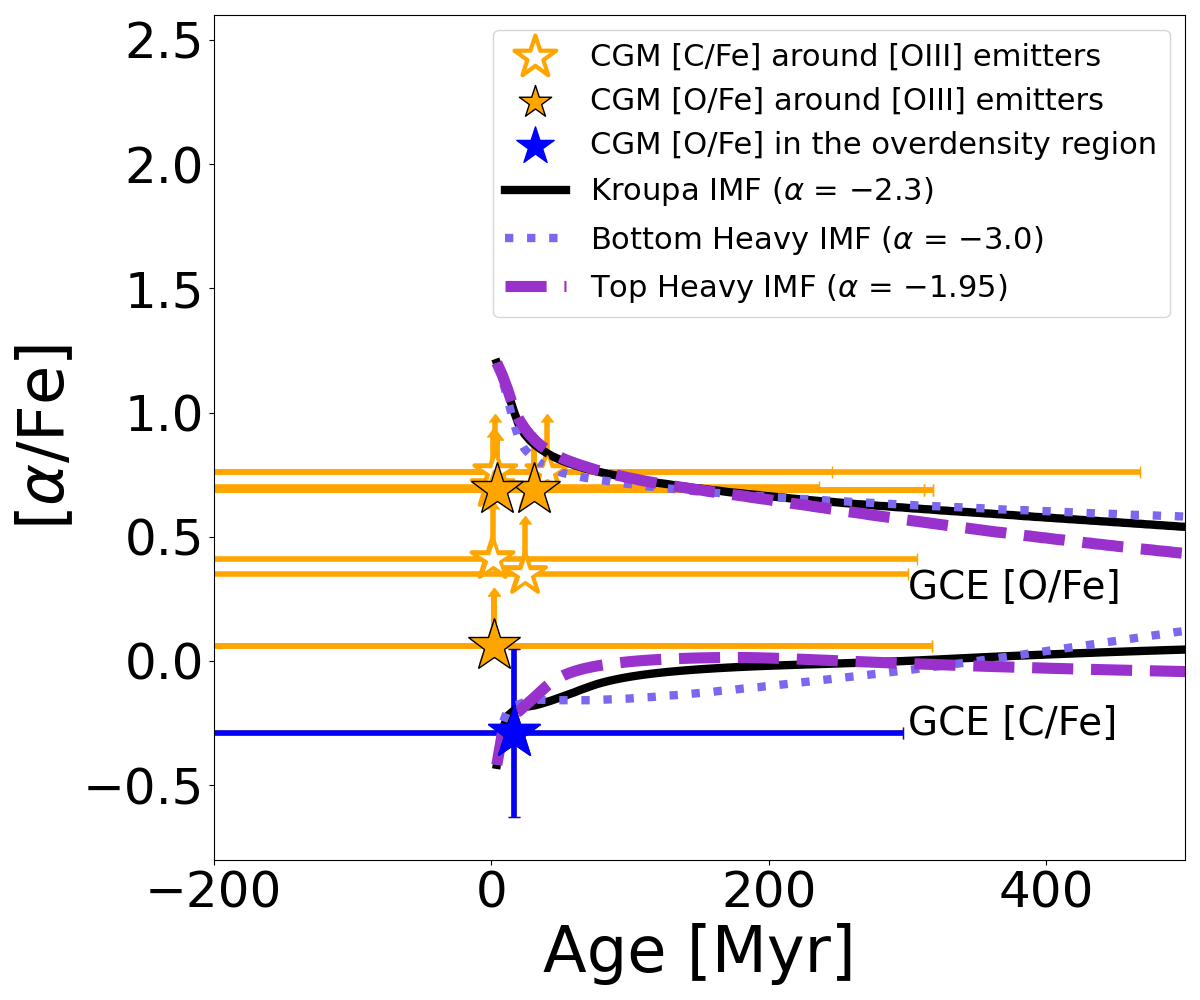}{0.5\textwidth}{}
\fig{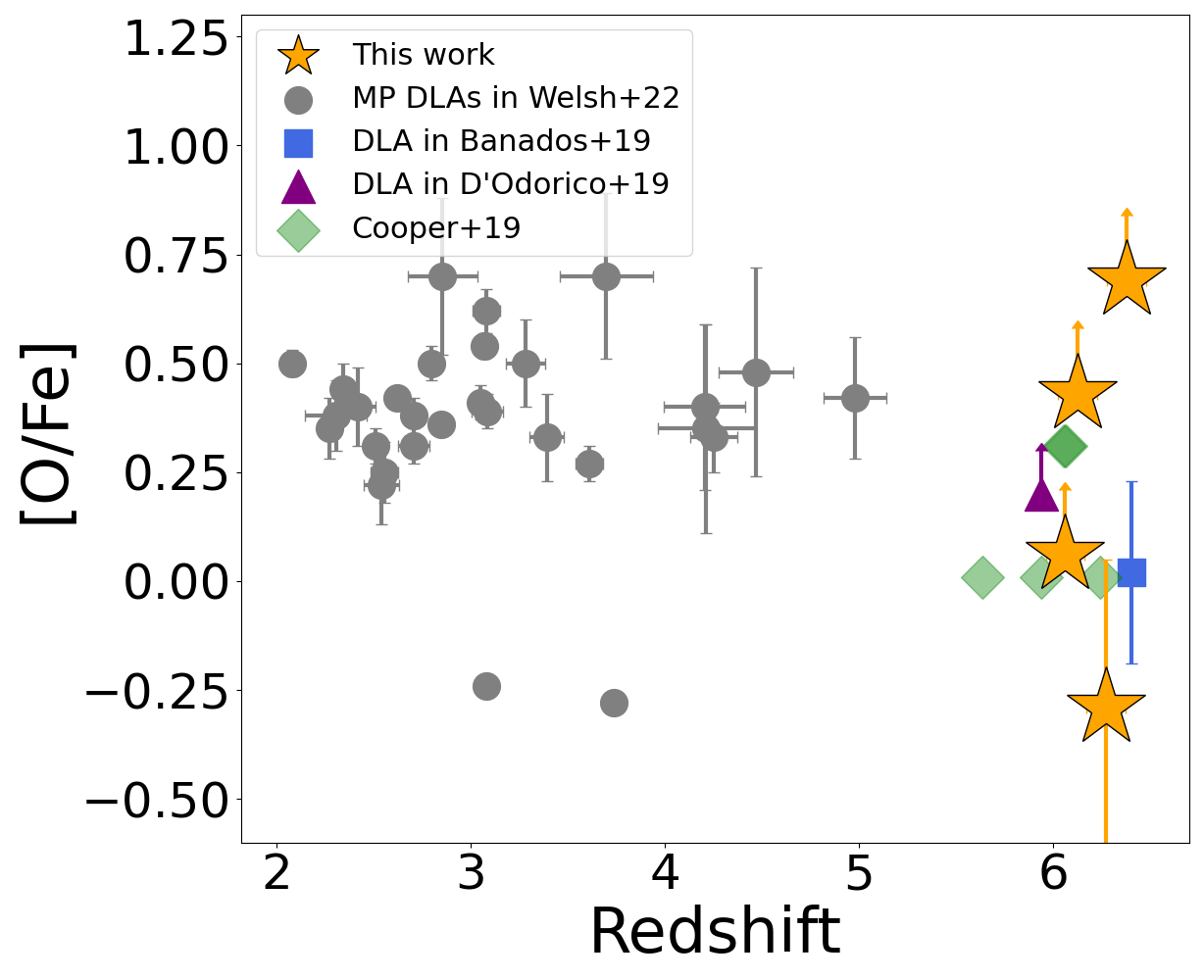}{0.5\textwidth}{}
}\caption{$Left$: Relative metal abundances between $\alpha$ (C and O) elements with iron (Fe) in the absorbing gas versus the age of associated galaxy candidates. The solid black, dashed blue and long dashed purple curve are the Kroupa, bottom heavy and top heavy IMF in the GCE models, respectively. $Right$: The comparison of [$\alpha$/Fe] ratio with that in metal-poor DLAs in \citet{welsh22} at $z<$ 5.7 and those detected at $z>$ 5.7.}\label{fig:alpha_fe}
\end{figure*}

 
In the left panel of Figure \ref{fig:alpha_fe}, we plot [C/Fe] and [O/Fe] values of the absorbing gas against the stellar age of their associated \oiii~emitter candidates. We find that the [$\alpha$/Fe] values for the CGM in a less dense region ($\leq$ three galaxies within 1 Mpc) are tentatively about 1 dex greater than that in the most overdense region (nine galaxies within 1 Mpc). This difference might hint at the galaxy assembly and star formation history during the cosmic dawn. We suggest two potential explanations for this observation: first, galaxy mergers or clumpy structures might reduce the production of $\alpha$ elements, particularly from Type-II SNe; or the interactions within galaxy groups might suppress the efficiency in expelling lighter elements like carbon and oxygen than Fe-peak elements into the CGM. Secondly, the overdensity region somehow trigger the Type-I SNe in generated Fe-peak elements earlier than expected. 

\subsection{Accuracy in relative abundance measurements}
Note that the accuracy of measurements on the gas-phase relative abundances would be affected by the potential line saturation and metal depletion onto the dust. From Appendix \ref{fig:vpfit-xqr30-a1}-\ref{fig:vpfit-xqr30-a4}, we can tell that for all the systems with detected low ions, there is no $>3\sigma$ Fe~{\sc ii} detection except for two systems ($z=$ 6.12279 and 6.2713). In those systems without Fe~{\sc ii} detection, the [O/Fe] or [C/Fe] values in Table \ref{table:absorber-galaxy} will be even higher if the O~{\sc i}, C~{\sc ii} and C~{\sc iv} lines are saturated/partially saturated.

The two systems having both O~{\sc i} and Fe~{\sc ii} detected show similar $N_{\text{O~{\sc ii}}}$. The Fe~{\sc ii} absorption detected at $z = 6.2713$ is more saturated than that at $z = 6.12279$. The [O/Fe] values are $0.56 \pm 0.22$ and $0.36 \pm 0.17$ with fixed $b$ values of 10 \kms~and 15.8 \kms(the $b$ value in \citealt{davies23a}) for the $z = 6.12279$ system. The [O/Fe] values are $-0.32 \pm 0.34$ and $-0.23 \pm 0.32$ with fixed $b$ values of 5 \kms~and 15 \kms~for the $z = 6.2713$ system. In summary, with careful consideration of line saturation, we confirm that our newly detected system in the J0305-3150 sightline has the lowest [O/Fe] value in this work.


Dust depletion may also affect the gas-phase relative metal abundance measurements. Different metal species can be locked into the dust grains with varying depletion factors. Iron has approximately 1 dex more depletion onto the dust than carbon and oxygen \citep{jens09}. The [C/Fe] and [O/Fe] will be slightly enhanced due to possible dust depletion of Fe. Nevertheless, at the metallicities assumed, Fe depletion would be very small and not significantly affect the results. Future ancillary ALMA observations and analyses in these quasar fields will be presented in the EREBUS collaboration (in prep).  


\subsection{Possibility of top-heavy IMF contribution to the CGM at the cosmic dawn}\label{sec:varying_imf}

In this section, we discuss the stellar population in the CGM host galaxy candidates and its effect on the metal abundances in the CGM. We note that several absorbing systems in our sample exhibit similar features as damped Lyman-$\alpha$ systems (DLAs) with the presence of singly ionized O~{\sc i} and Mg~{\sc i} absorption. 

\citet{girish13} perform a self-consistent model on the Population III stars contribution in the IGM at the reionization epoch. They found that the chemical signatures of Population III stars remain in low-mass galaxies (halo mass $<10^9$ M$_\odot$) at $z\sim$ 6, and the relative abundances of the metal-poor DLAs are likely holds the promise to constrain Population III enrichment in the early universe. \citet{welsh22} model the contributions from both Population II and Population III stars in metal-poor DLAs. \citet{sac23} studied the metal abundances in 37 LLS and sub-DLAs at $z\sim$ 3 -- 4.5 and found that among the 14 very metal-poor LLSs or sub-DLAs in their sample, the [C/Fe] and [O/Fe] are similar to those in galactic metal-poor stars and ultra-faint dwarf galaxies.

In the right panel of Figure \ref{fig:alpha_fe}, we compare the relative abundances, [O/Fe], of the four systems with clear O~{\sc i} detection in our sample to those detected in metal-poor DLAs to date and two DLAs detected at $z \sim 5.9$ \citep{dod18} and $z \sim$ 6.4 \citep{bana19} and low-ion absorbers in \citet{coo19}.
We note that our observed [O/Fe] is similarly to that in metal-poor DLAs and LLSs at lower redshift, possibly suggesting a pristine environment to form the early generation of stars. If these systems are DLAs, their metallicities are plausibly close to the average DLA metallicity at $z\sim$ 6 as predicted by the empirical relation in \citet{raf12}: $<Z>$ = (--0.22$\pm$0.05)$\times z$ -- (0.66$\pm$0.15), i.e., --1.95$\pm$0.45 at $z =$ 6.

We further consider the effects of different IMF on the metal abundances of dwarf galaxies at the cosmic dawn and thus the metal abundances in the CGM/IGM. IMFs are generally modeled as a power law of index $\alpha$. We updated the galactic chemical evolution (GCE) model in \citet{cote17} and the algorithm NuPyCEE\footnote{https://github.com/NuGrid/NuPyCEE} by varying the stellar IMF. Details of the updated algorithm and model assumption are presented in Guo et al. (in prep). We take the galaxy's stellar mass, SFR, and age from Table \ref{table:absorber-galaxy} as the model inputs. The outputs are the stellar yields of different metals. Specifically, we set three IMFs: top-heavy IMF ($\alpha$ = --1.95), Kroupa IMF ($\alpha$ = --2.3), and bottom-heavy IMF ($\alpha$ = --3.0), with stellar mass ranging between 1--40 $M_{\odot}$. We plot the modeling results in the left panel of Figure \ref{fig:alpha_fe}. We find that for an isolated dwarf galaxy at $z\sim$ 6, a top-heavy IMF generates three times higher [C/Fe] and similar [O/Fe] than the Kroupa IMF does when the stellar age is smaller than 200 Myr.  If the stellar feedback occurs simultaneously with the star formation and blows the stellar yields into its CGM/IGM, we may detect similar [C/Fe] overabundance in its CGM/IGM. Our GCE modeling in dwarf galaxies at $z\sim$ 6 can account for the observed CGM [O/Fe]. The [C/Fe] in the CGM favors a top-heavy IMF in the associated galaxies. However, the top-heavy IMF alone in our GCE model cannot fully account for the detected carbon overabundance in the CGM. The top-heavy IMF can only reach C/Fe values of 0.015, which is smaller than our lowest measured value of 0.06. We note that the [C/Fe] yields PopIII stars in the model of \citet{girish13} can potentially explain our measured [C/Fe] values. This may indicate that our detected CGM gas could have been polluted by the PopIII stars in nearby galaxies.

\subsection{Surrounding galaxies as local ionizing sources of the CGM}\label{sec:cloudy}



\begin{figure*}[htb!]
\gridline{
\fig{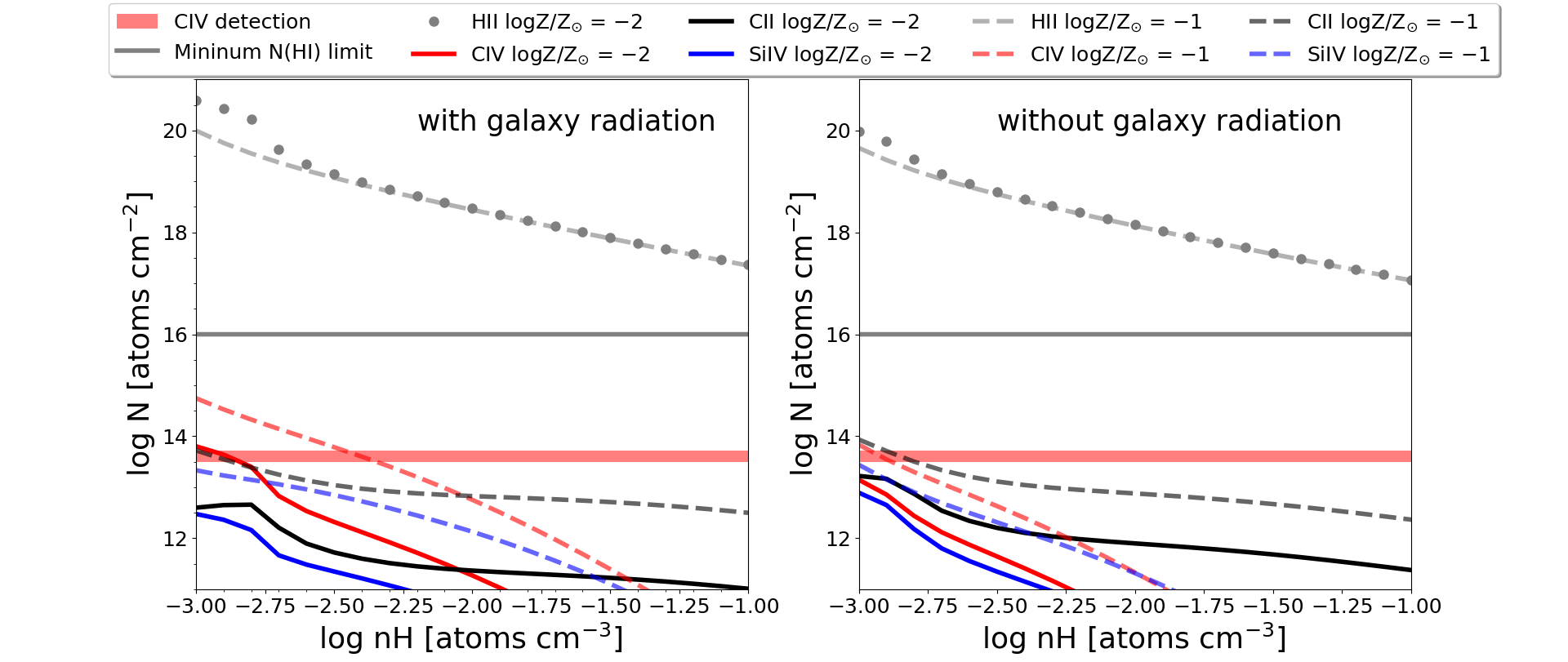}{1.0\textwidth}{}
}
\caption{Photoionization modeling of the absorbing gas in the absorber system at $z$ = 6.17255 towards J0224--4711 using CLOUDY. The left and right panels are the models with and without the UV flux from the detected nearby galaxies, respectively. We can tell that without an external UV flux, the metagalactic UVB background only cannot reproduce the observed [C~{\sc iv}/C~{\sc ii}] ratio.
}\label{fig:cloudy}
\end{figure*}

\citet{sch06} point out that the local ionizing source has a significant effect on the ionization state of the IGM gas. \citet{fin16} examined the effects of different UV background (UVB) sources on the strengths of ions (C~{\sc iv}, C~{\sc ii}, and Si~{\sc iv}) in the CGM/IGM. They found that the UVB from \citet{hm12} only overpredicts the C~{\sc ii}/C~{\sc iv} abundance in highly-ionized C~{\sc iv} systems, suggesting a local amplification of UV radiation from galaxies. Their simulations also suggest that the strong C~{\sc iv}-galaxy correlation extends to at least 300 pkpc.



We thus perform a simple test on the ionizing radiation from our detected absorber-associated galaxy candidates and its effect on the CGM gas ionization state, in particular, the systems with only C~{\sc iv} systems detected (log $N_{\text{C~{\sc iv}}}>$ 13.0). First, we estimate the ionizing radiation bubble size generated by the \oiii~emitters. 
We calculate the galaxy radiation radius using its Str\"omgren radius $(R_s)$ from the analytical model described in \citet{bolton07a}. The Str\"omgren radius is defined by

\begin{eqnarray}
R_s &=& \left( \frac{3N_{ion}}{4\pi \alpha_{rec}n^2} \right)^{1/3} \nonumber \\
&\sim& 5\textrm{Mpc} \left( \frac{N_{ion}}{10^{55}} \right) \times \left( \frac{n}{n_{H_0}}\Delta \right)^{-2/3} \times f_{esc}^{1/3}
\end{eqnarray}

Two free parameters are in this model: the gas overdensity $\Delta= \frac{n}{n_{H_0}}$ and the galaxy escape fraction $f_{esc}$. The $f_{esc}$ of galaxy at $z\geq$ 6 has a large variety (0.01\%–20\%, \citealt{ma15}). If an \oiii-emitting galaxy has $M_{1450}$ = --20 (luminosity $L_{1450}\sim$ 10$^{45}$ erg/s) and if we assume its $f_{esc}$ = 0.1 and adopt the gas density $n_H$ = 10$^{-3}$ cm$^{-3}$ from our CLOUDY modelling, its Str\"omgren radius $R_s$ is $\sim$ 1 Mpc. The H~{\sc ii} expansion rate can be calculated following the relation $R(t) = R_s\times(1-e^{-t/t_r})^{1/3}$, where the $t_r$ = 1/($n_H\alpha$) and $\alpha$ is the recombination rate (2.6$\times 10^{-13}$ cm$^{-3}$ s$^{-1}$). Then the timescale of $R_s$ expands to 300 pkpc is around 3 Myr, which is reasonable within our galaxy age.

We then conduct photoionization modeling to obtain the physical properties of the absorbing gas using \textsc{CLOUDY} \citep{cloudy17}. In the models, we include the metagalactic UV background from \citet{khai19}, the cosmic microwave background at \( z = z_{\text{abs}} \), and cosmic rays as the extragalactic UV background (UVB). The UV flux from the detected \oiii~emitters is estimated from its UV magnitude $M_{UV}$ from our SED fitting. For the neutral systems ($N_{\text{C{\sc ii}}}/N_{\text{C{\sc iv}}}>$ 3), we find that the observed $N_{\text{C{\sc ii}}}/N_{\text{C{\sc iv}}}$ can be reproduced without adding a local UV flux. For the highly ionized C~{\sc iv} gas ($N_{\text{C{\sc ii}}}/N_{\text{C{\sc iv}}}<$ 1), an additional ionizing source is required to reproduce the C~{\sc iv}/C~{\sc ii} ratio. We show the fitting result of the \( z = 6.17255 \) system towards J0224M4711 in Figure \ref{fig:cloudy}.


Direct detections of galaxies around highly ionized C~{\sc iv} systems ($N_{\text{C{\sc ii}}}/N_{\text{C{\sc iv}}}<$ 3) at $z>$ 6 are rare (see \citealt{diaz11, diaz15, mey19} for connections between C~{\sc iv} and \lya~emission). The galaxy around the highly ionized C~{\sc iv} system (log $N_{\text{C{\sc iv}}}$ $\sim$ 13.4) at 5.9784 in \citet{diaz11} has $M_{UV}$ of --20.66$\pm$0.05 and $D$ = 311.4 pkpc. The galaxy detected around the $z=$ 5.7242 C~{\sc iv} system (log $N_{\text{C{\sc ii}}}$ $\sim$ 14.52) in \citep{diaz15} has $M_{UV}$ = --20.65$\pm$0.52 and $D$ = 212.3 pkpc. \citet{mey19} statistically calculated the correlation between \lya~emission around C~{\sc iv} systems at 5.7 $<z<$ 6.2 and concluded that C~{\sc iv} absorbers with log $N_{\text{C~{\sc iv}}}>$ 13.2 are associated with galaxies having $M_{UV}<$ --16.  The C~{\sc iv}-only system at $z$ = 6.4821 that we do not find any galaxy within 1 pMpc has the lowest $N_{\text{C{\sc iv}}}$, may associated with galaxies fainter than $M_{UV}=$ --16.0. In summary, our detection of highly ionized C~{\sc iv} gas [O~{\sc iii}]-emitting galaxies and their effects on the local UVB are consistent with cosmological simulations and previous limited detections.

\section{Summary}
1. We detected a new absorber at $z$ = 6.2713 from quasar sightline J0305--3150. Along with absorbers from \citet{davies23a}, we have a sample of nine spanning $z$ = 6.03 to $z$ 6.49. Using ASPIRE JWST/WFSS data, we found 8 (11) \oiii-emitters within $D$ = 350 (1 Mpc) of these absorbers, with stellar masses log \(M_*/M_{\odot}\) from 7.2 to 8.8, and metallicities \(Z/Z_{\odot}\) from 0.02 to 0.4 solar metallicity.

2. We find that the absorbing gas with a higher [$\alpha$/Fe] is possibly associated a less overdense region than the most overdense region. 

3. We examine various IMFs in galaxies at $z\sim$ 6 with ages less than 500 Myr. Our findings indicate that the top-heavy IMF produces a [C/Fe] that is three times higher than the Kroupa IMF \citep{kroupa01} during the initial 200 Myr. The yields from the Pop III IMF suggest its potential contribution to the [C/Fe] overabundance in our observed CGM. 


4. We perform photoionization modeling for the all the systems and find that a local ionizing source is plausible for the highly ionized C~{\sc iv} systems, {which is consistent with the cosmological simulation results at $z\sim$ 6.}

5. We note that fainter galaxies below our detection limit may reside closer to our absorbing gas in the CGM, this will not affect our discussion on the CGM metal abundance and its ionization source. 


\clearpage

\acknowledgments
We acknowledge the very constructive comments provided by the anonymous referee. We appreciate fruitful discussions with Renyue Cen, Xiaoting Fu, Zhiyu Zhang, Fuyan Bian, Alba V. Garcia, Michele Fumagalli, and Patrick Petitjean. Furthermore, we would like to thank the public spectra and catalog from the XQR-30 team. SZ, ZC, XL, ZL, YW and ML acknowledge support from the National Key R\&D Program of China (grant no.\ 2018YFA0404503), the National Science Foundation of China (grant no.\ 12073014, no.\ 12303011). The science research grants from the China Manned Space Project with No. CMS-CSST2021-A05, and Tsinghua University Initiative Scientific Research Program (No. 20223080023). SEIB is funded by the Deutsche Forschungsgemeinschaft (DFG) under Emmy Noether grant number BO 5771/1-1. KI acknowledge support from the National Natural Science Foundation of China (12073003, 11991052, 11721303, 11950410493), and the China Manned Space Project Nos. CMS-CSST-2021-A04 and CMS-CSST-2021-A06. GK is partly supported by the Department of Atomic Energy (Government of India) research project with Project Identification Number RTI$\sim$4002, and by the Max Planck Society through a Max Planck Partner Group. This work is based on observations made with the NASA/ESA/CSA James Webb Space Telescope. The data were obtained from the Mikulski Archive for Space Telescopes at the Space Telescope Science Institute, which is operated by the Association of Universities for Research in Astronomy, Inc., under NASA contract NAS 5-03127 for JWST. These observations are associated with program \#2078. Support for program \#2078 was provided by NASA through a grant from the Space Telescope Science Institute, which is operated by the Association of Universities for Research in Astronomy, Inc., under NASA contract NAS 5-03127. Some of the data presented in this paper were obtained from the Mikulski Archive for Space Telescopes (MAST) at the Space Telescope Science Institute. The specific observations analyzed can be accessed via \dataset[DOI: 10.17909/vt74-kd84]{https://doi.org/10.17909/vt74-kd84}.

\facilities{JWST (NIRCam), VLT (X-shooter), HST (ACS, WFC3)}







\bibliographystyle{aasjournal}
\bibliography{./ms}


\appendix
\restartappendixnumbering
\section{Figures}

\begin{figure*}
\gridline{
\fig{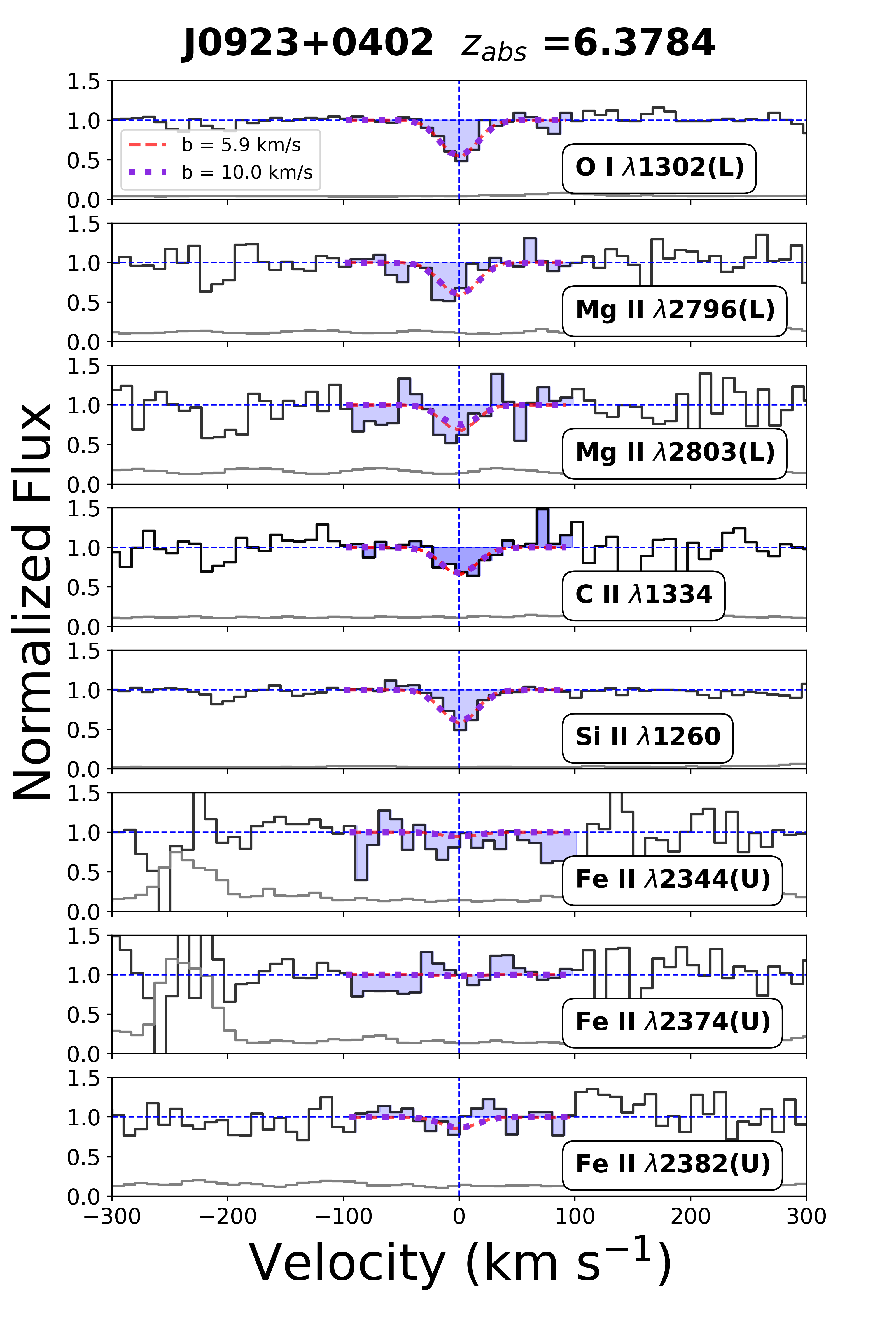}{0.5\textwidth}{}
\fig{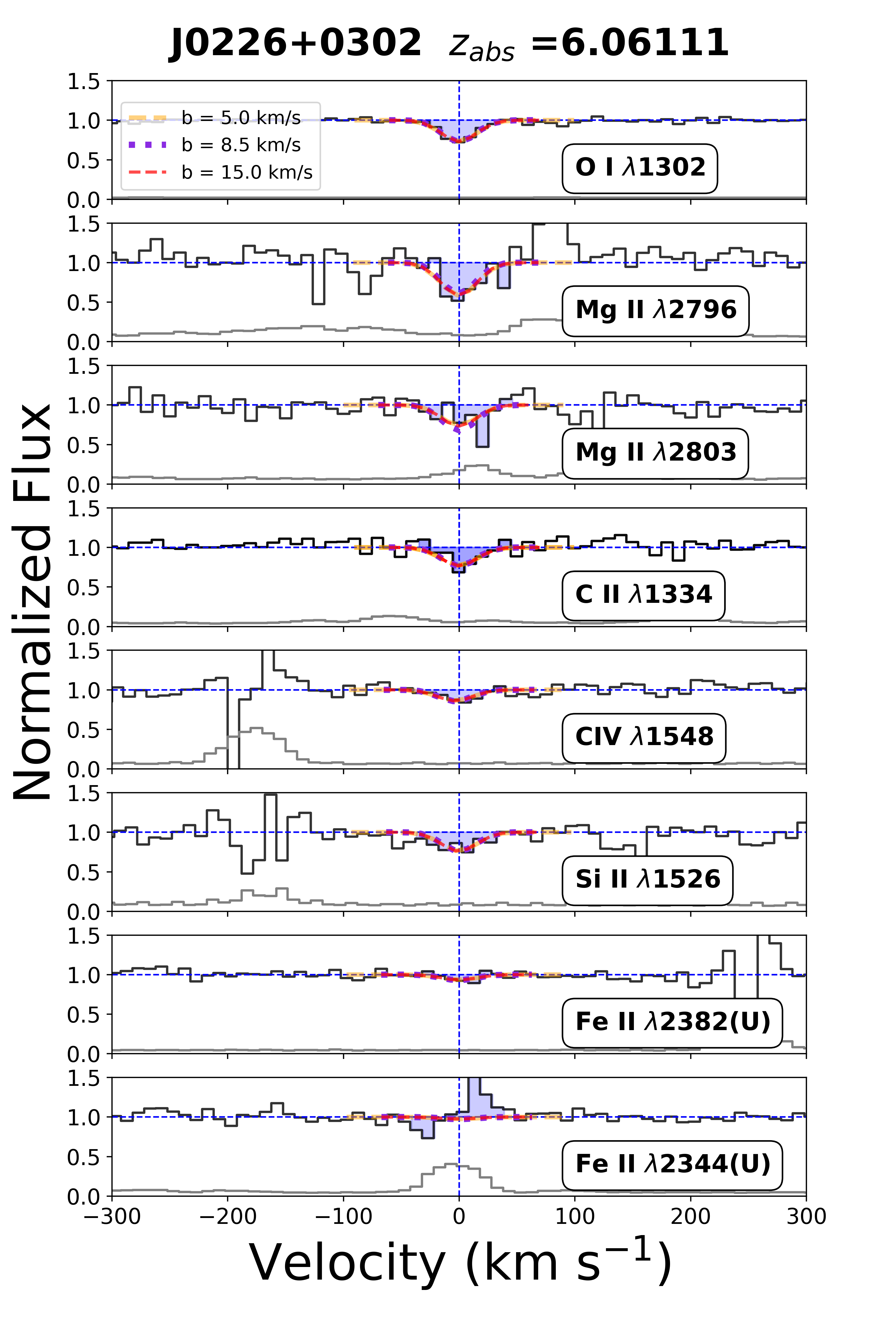}{0.5\textwidth}{}
}
\caption{The fitting of the metal lines with a Voight profile in the absorbing system at $z$ 6.3784 towards quasar J0923+0402 (left) and 6.06111 towards quasar J0226+0302. The red, purple and orange curves are the Voight fitting profile with different Doppler parameters. The red curve is the $b$ value used in \citet{davies23a}. The letters U and L represent the upper and lower limits on the column density measurement of the lines.}\label{fig:vpfit-xqr30-a1}
\end{figure*}

\begin{figure*}[htb!]
\gridline{
\fig{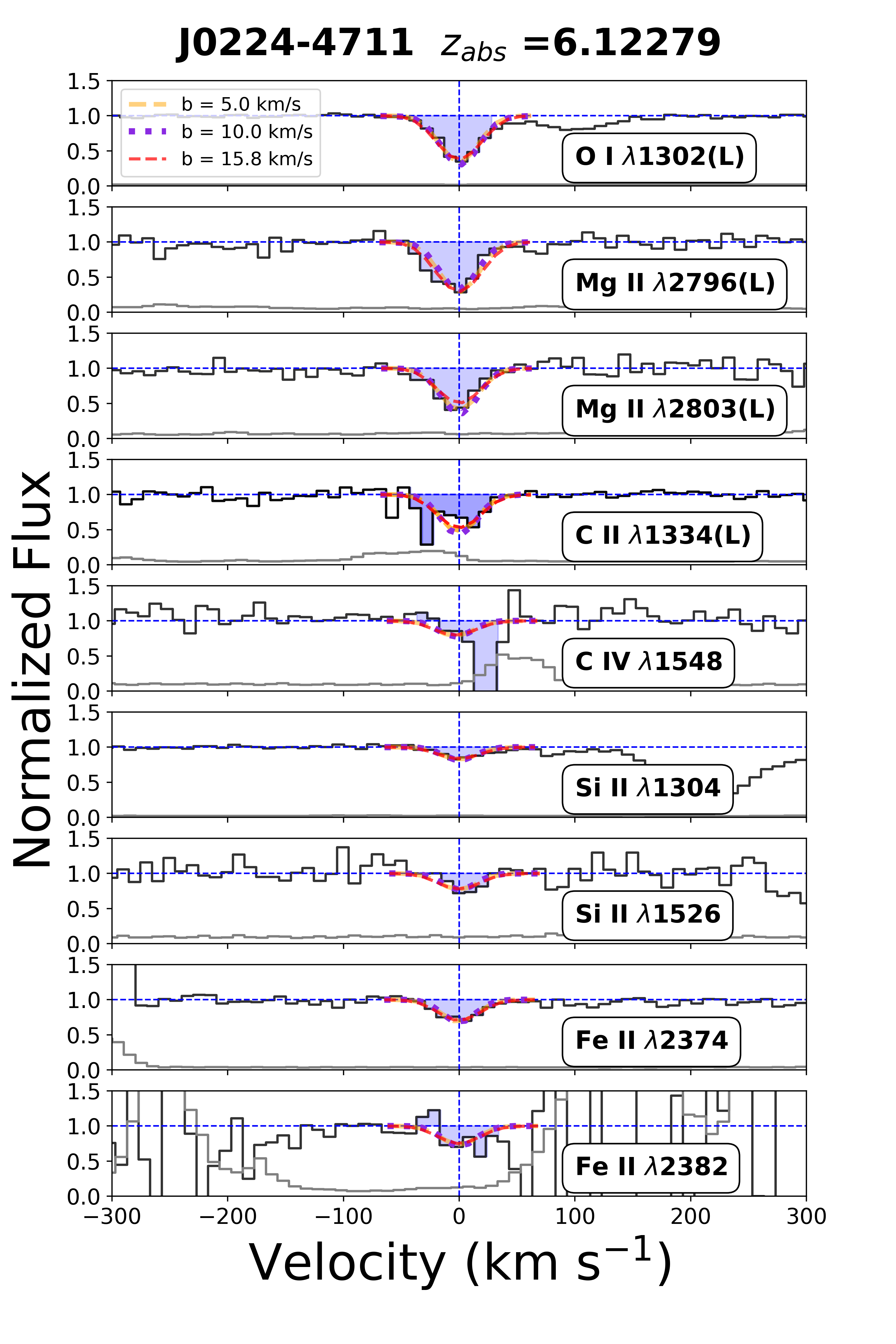}{0.5\textwidth}{}
\fig{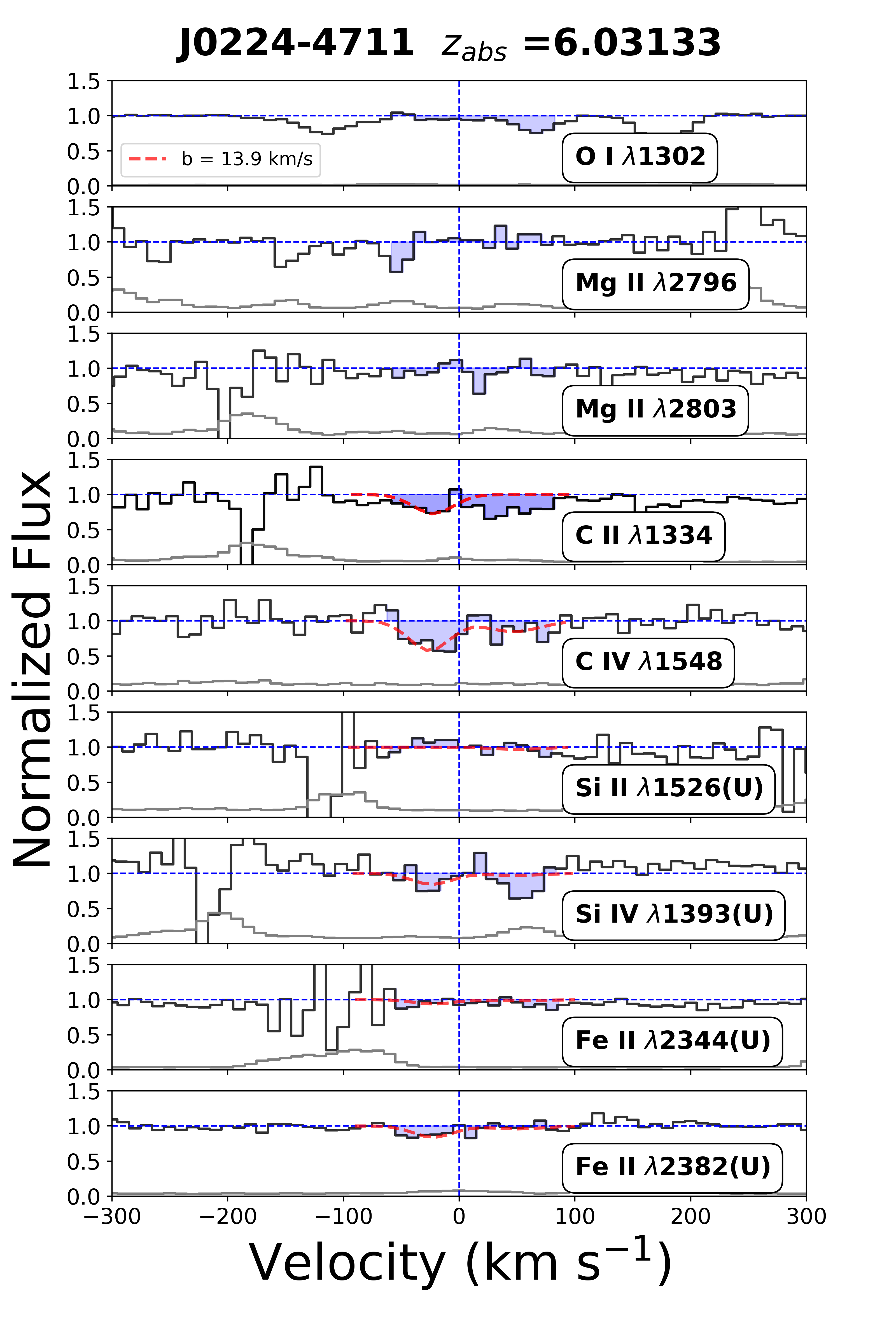}{0.5\textwidth}{}
 }
 \caption{The fitting of the metal lines with a Voight profile in the absorbing system at $z$ 6.12279 and $z$ = 6.03133 towards quasar J0224-4711. The red, purple and orange curves are the Voight fitting profile with different Doppler parameters. The red curve is the $b$ value used in \citet{davies23a}. The letters U and L represent the upper and lower limits on the column density measurement of the lines.}\label{fig:vpfit-xqr30-a2}
 \end{figure*}
  
\begin{figure*}
\gridline{
\fig{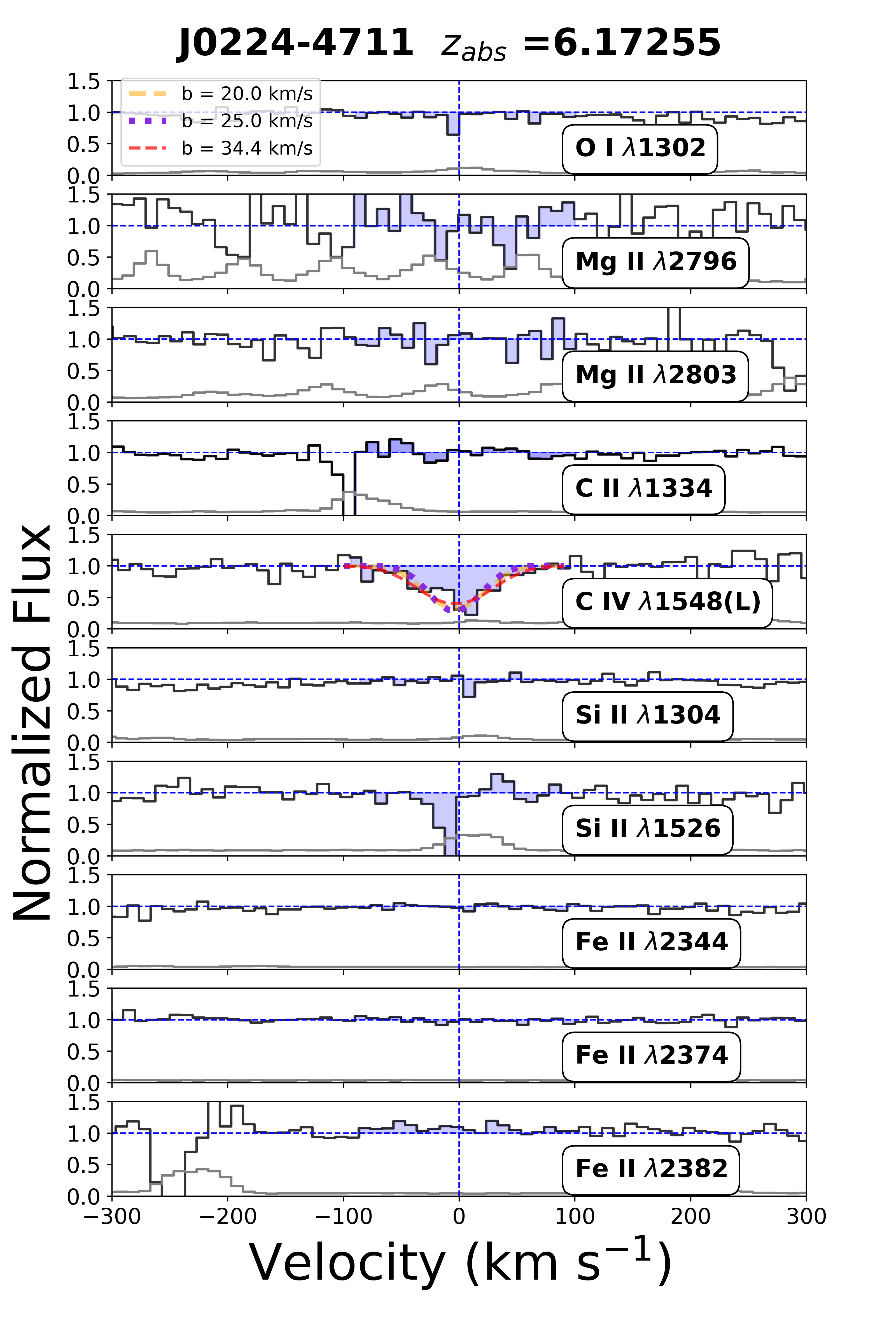}{0.5\textwidth}{}
\fig{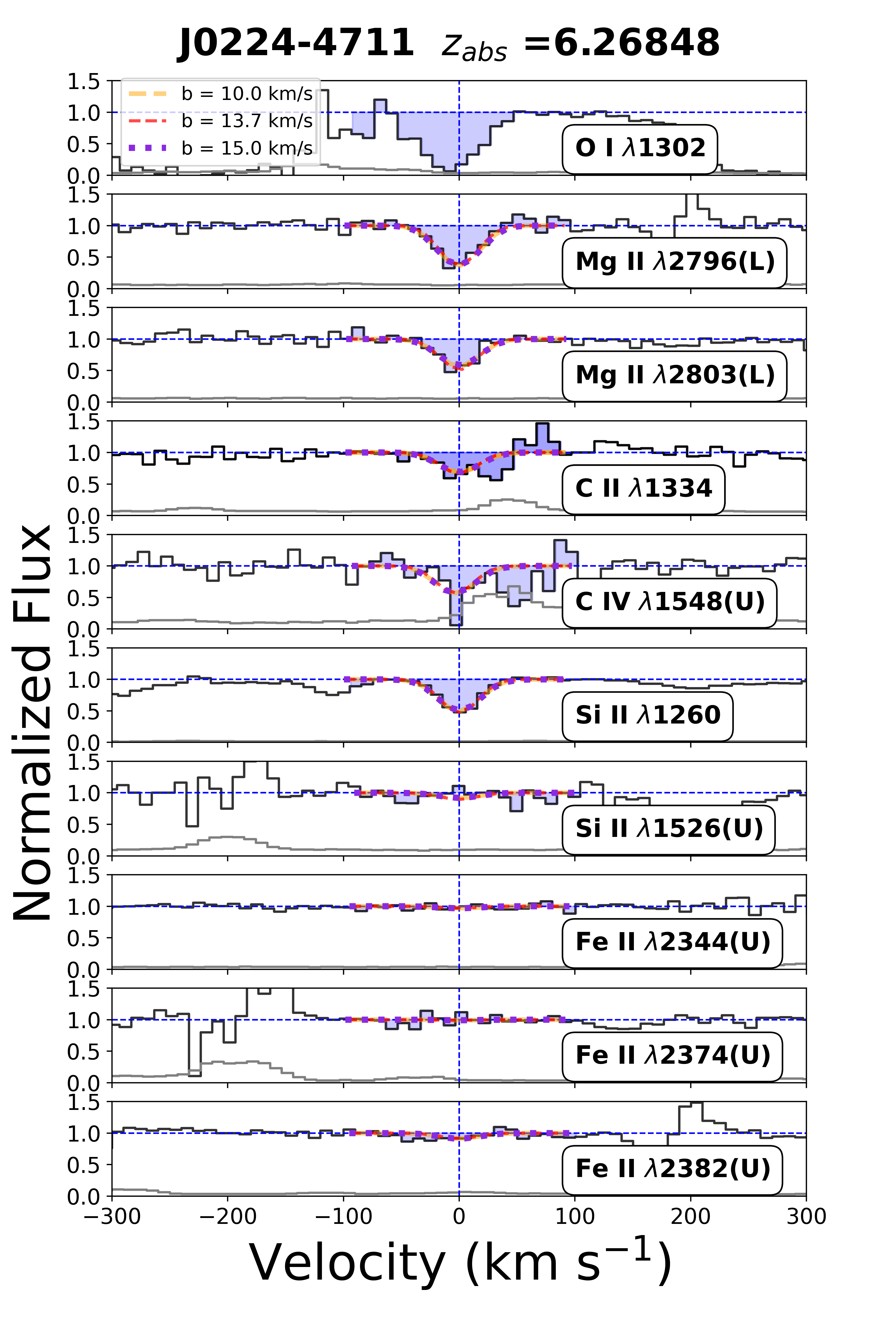}{0.5\textwidth}{}
}

\caption{The fitting of the metal lines with a Voight profile in the absorbing system $z$ = 6.17255 (left) and 6.26848 (right) towards quasar J0224-4711. The O~{\sc i}$\lambda$1302 and Si~{\sc ii}$\lambda$1304 lines are strongly blended with the C~{\sc iv} doublet at $z$ = 5.10911, the Si~{\sc iv} doublet are strongly blended with the sky lines, therefore, we do not plot in the right panel. The red, purple and orange curves are the Voight fitting profile with different Doppler parameters. The red curve is the $b$ value used in \citet{davies23a}. The letters U and L represent the upper and lower limits on the column density measurement of the lines.}\label{fig:vpfit-xqr30-a3}
\end{figure*}

\begin{figure*}
\gridline{
\fig{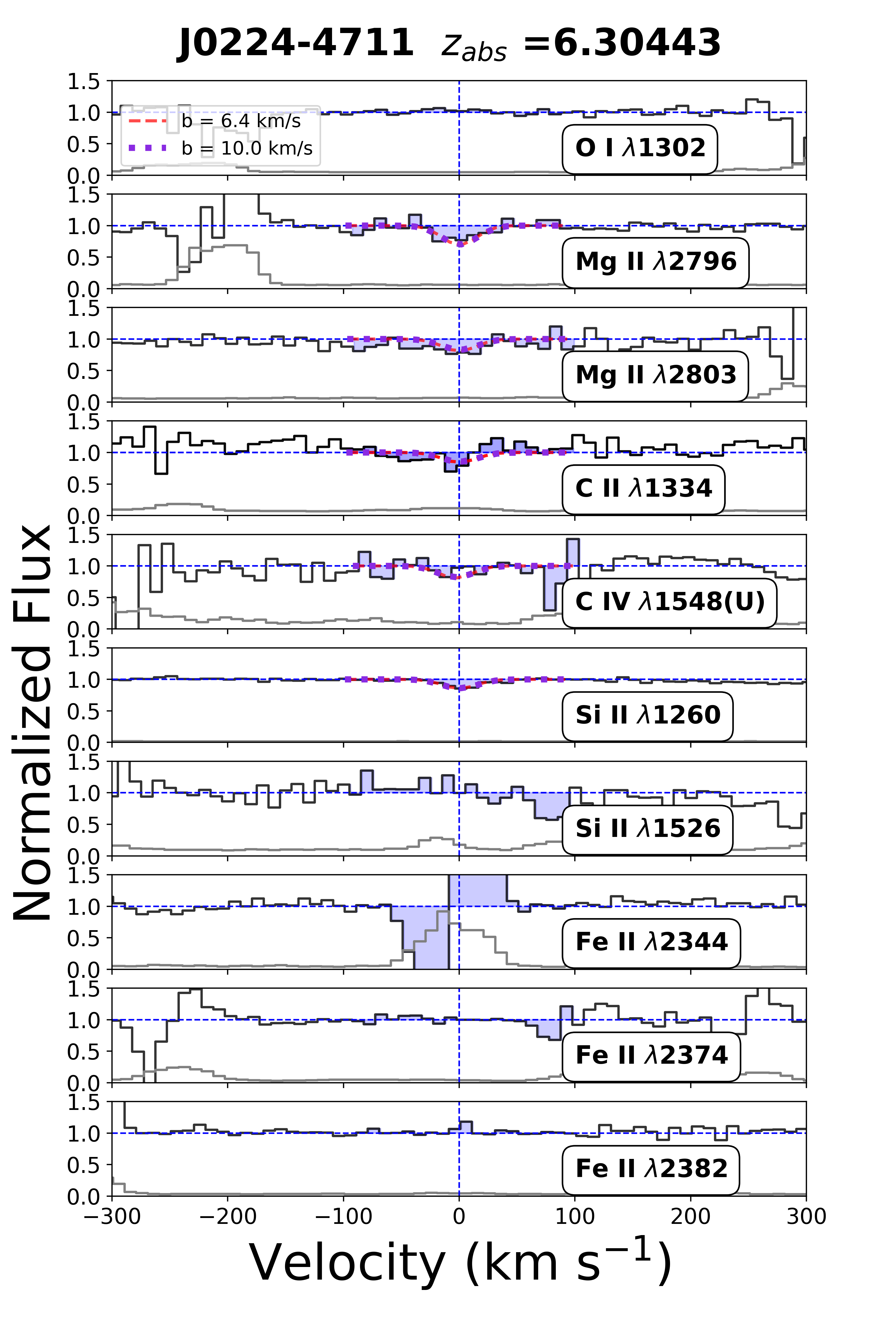}{0.5\textwidth}{}
}

\caption{The fitting of the metal lines with a Voight profile in the absorbing system $z$ = 6.30443 towards quasar J0224-4711. Weak Mg~{\sc ii}, C~{\sc ii}, and Mg~{\sc ii}($\lambda$1260) lines is detected. The letters U and L represent the upper and lower limits on the column density measurement of the lines.}\label{fig:vpfit-xqr30-a4}
\end{figure*}

\begin{figure*}[htb!]
\gridline{
\fig{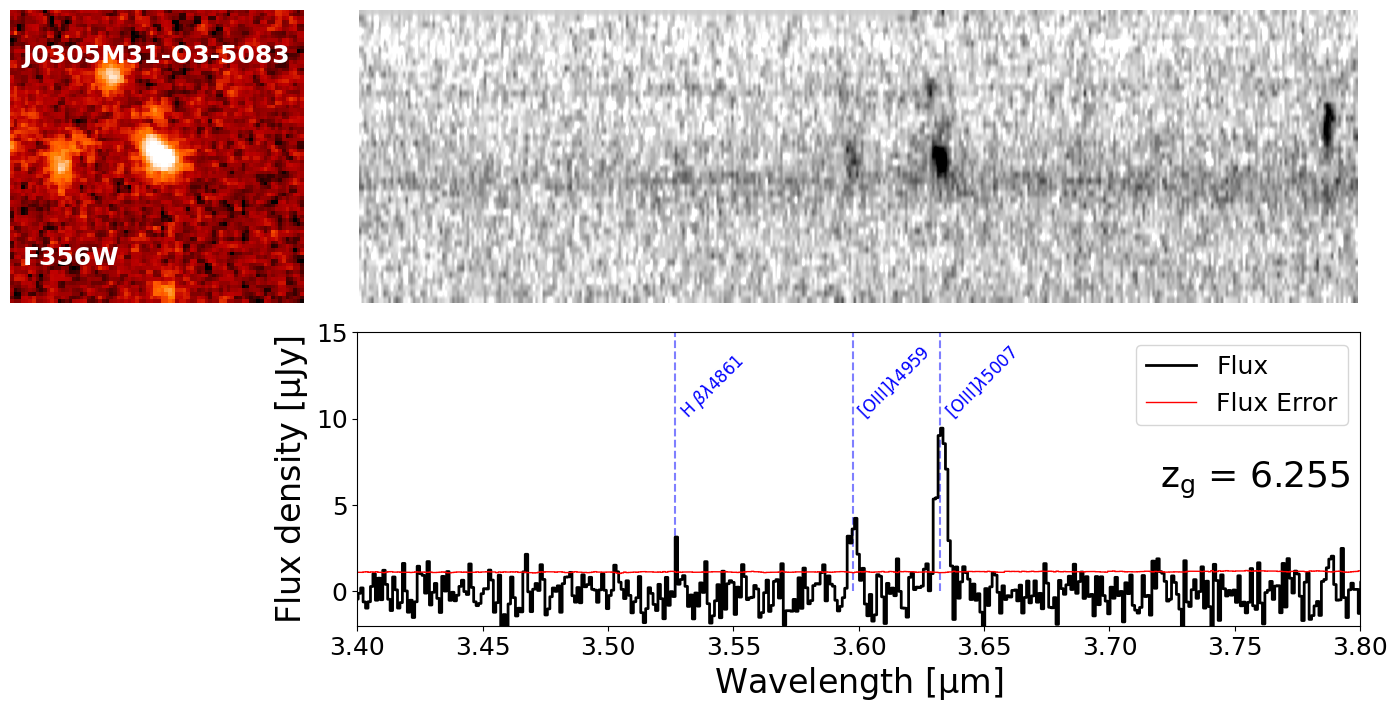}{0.5\textwidth}{}
\fig{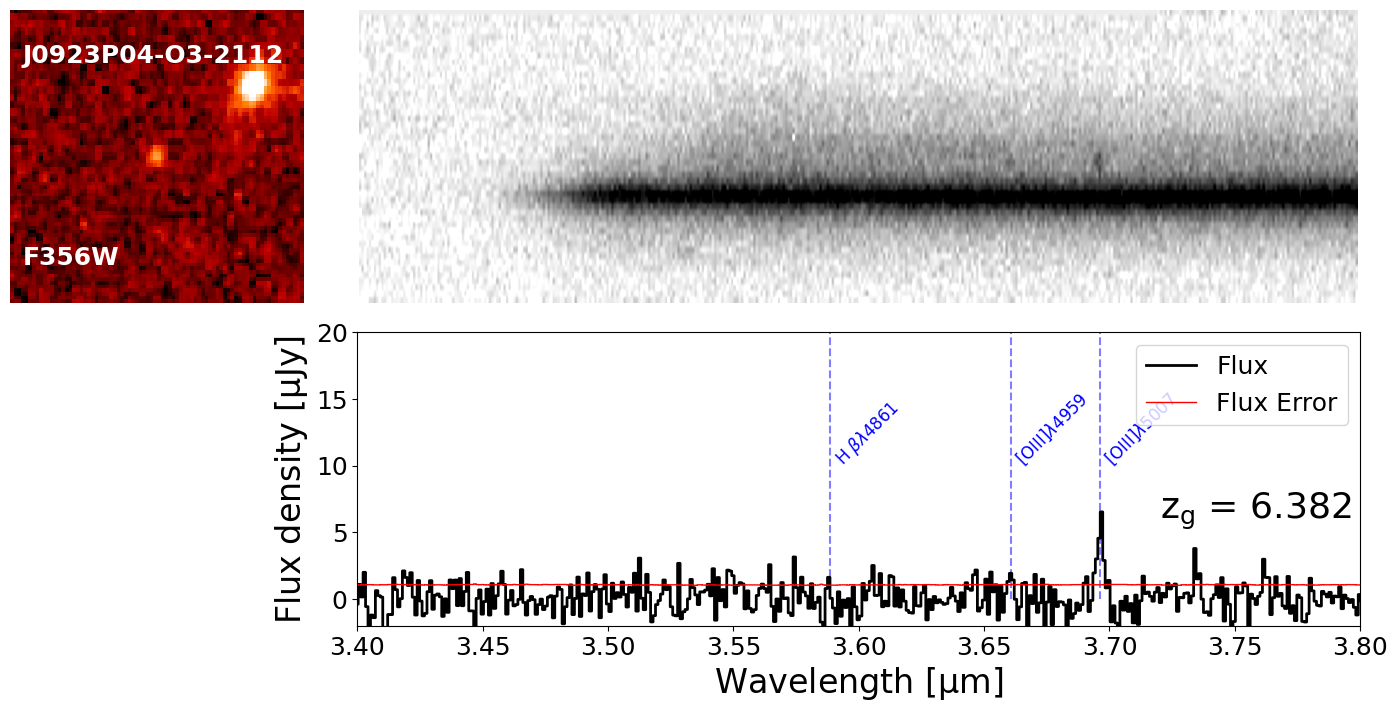}{0.5\textwidth}{}
}

\gridline{
\fig{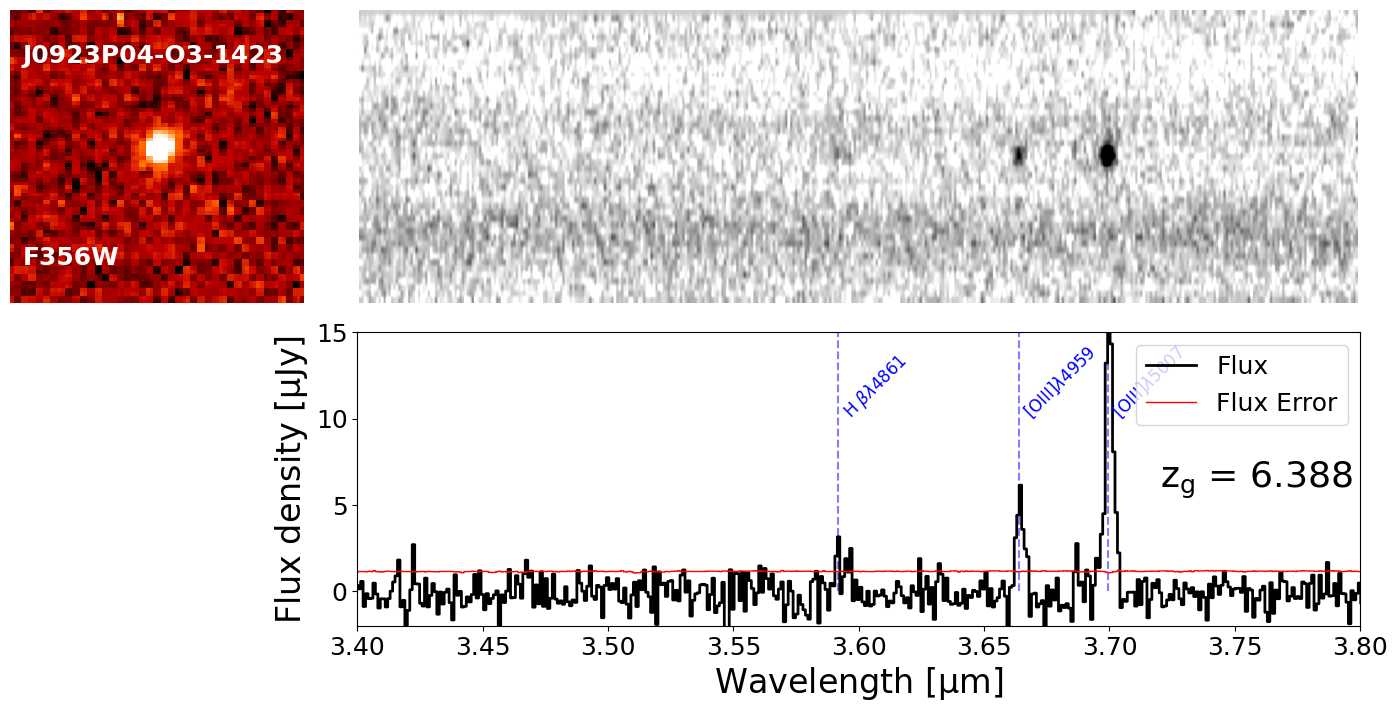}{0.5\textwidth}{}
\fig{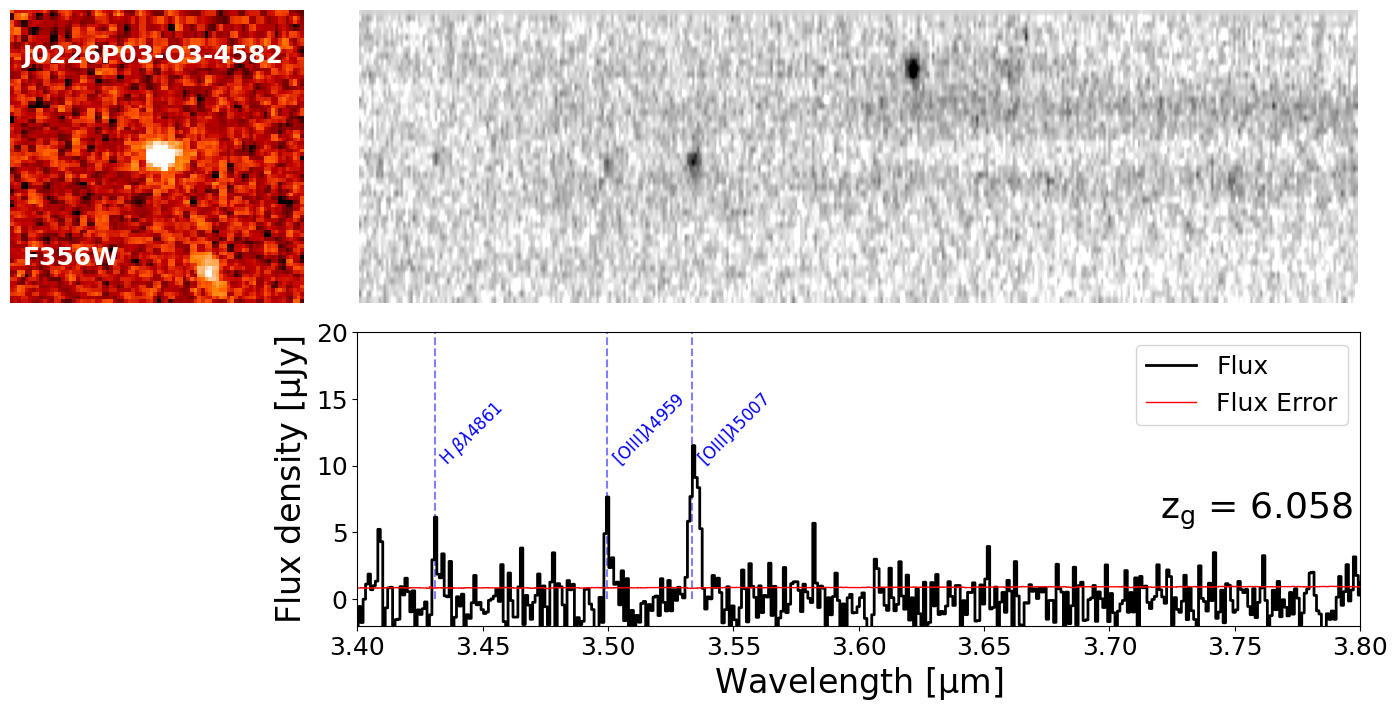}{0.5\textwidth}{}
}

\gridline{
\fig{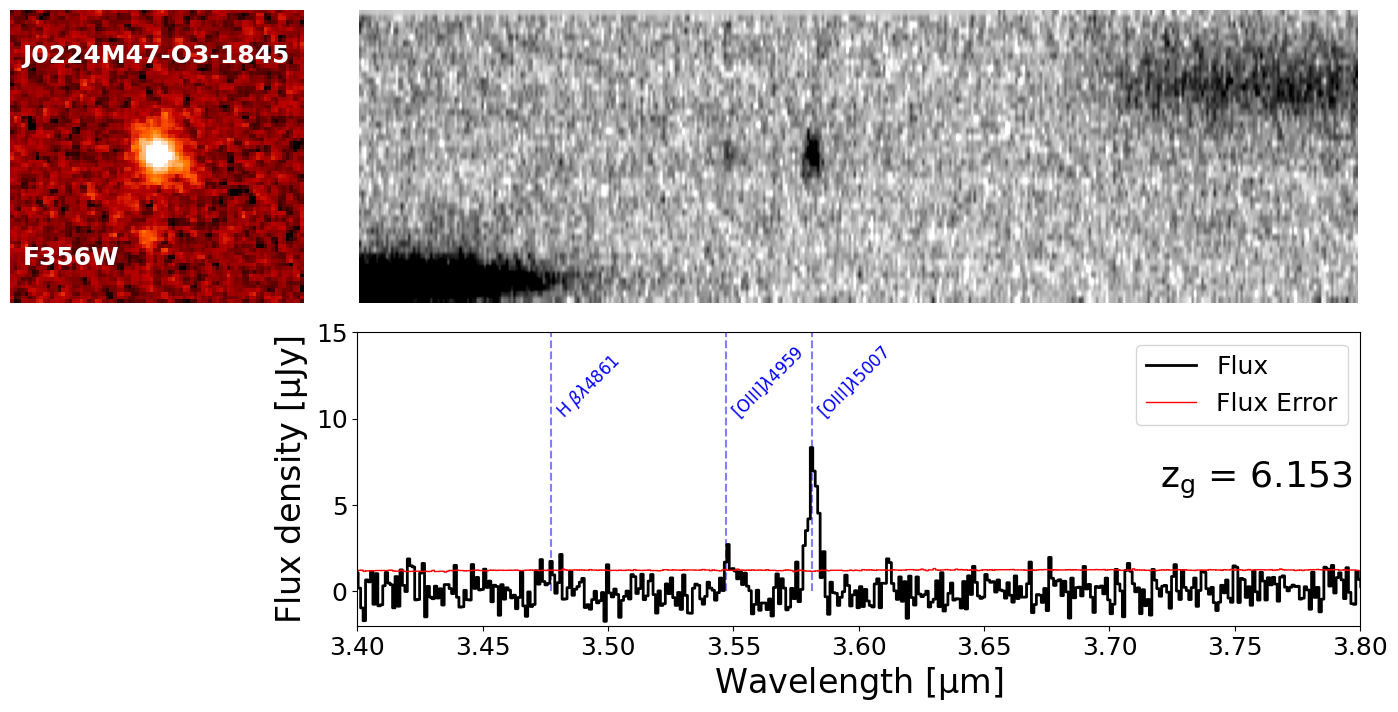}{0.5\textwidth}{}
\fig{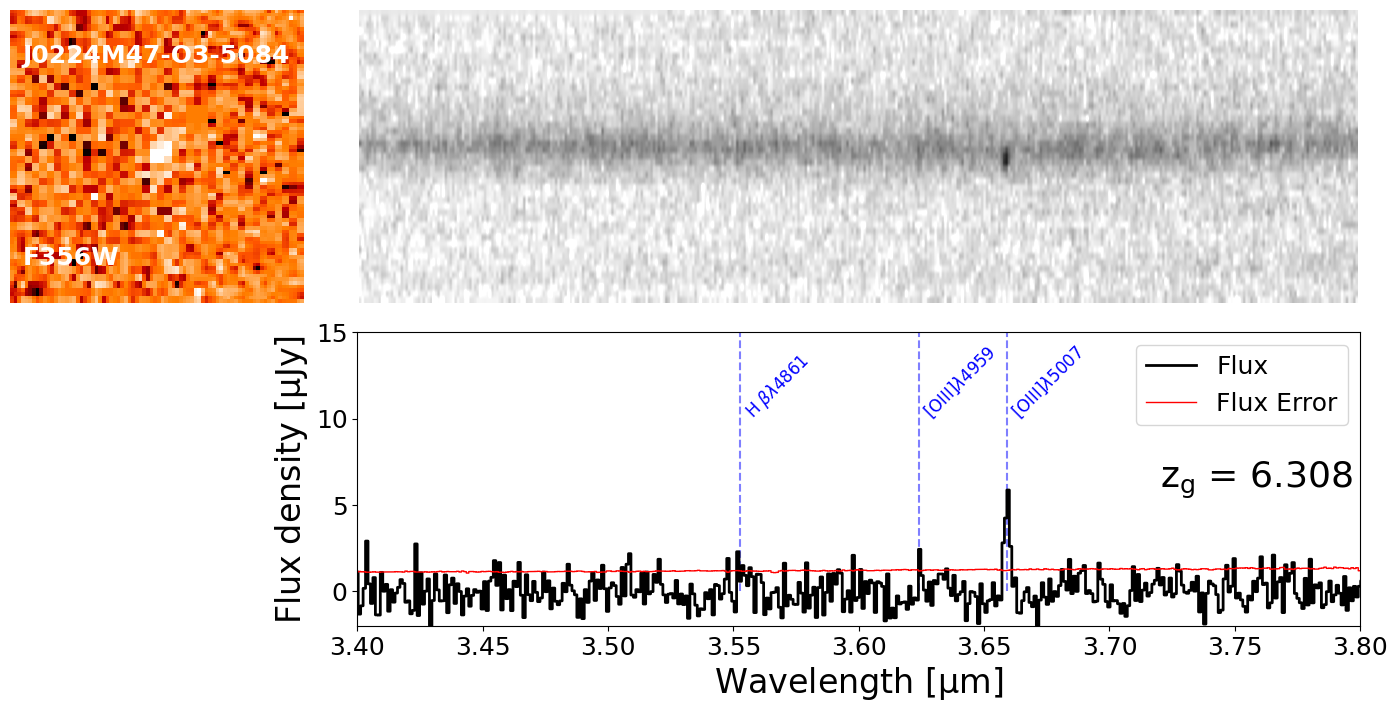}{0.5\textwidth}{}
}

\end{figure*}

\begin{figure*}[htb!]

\gridline{
\fig{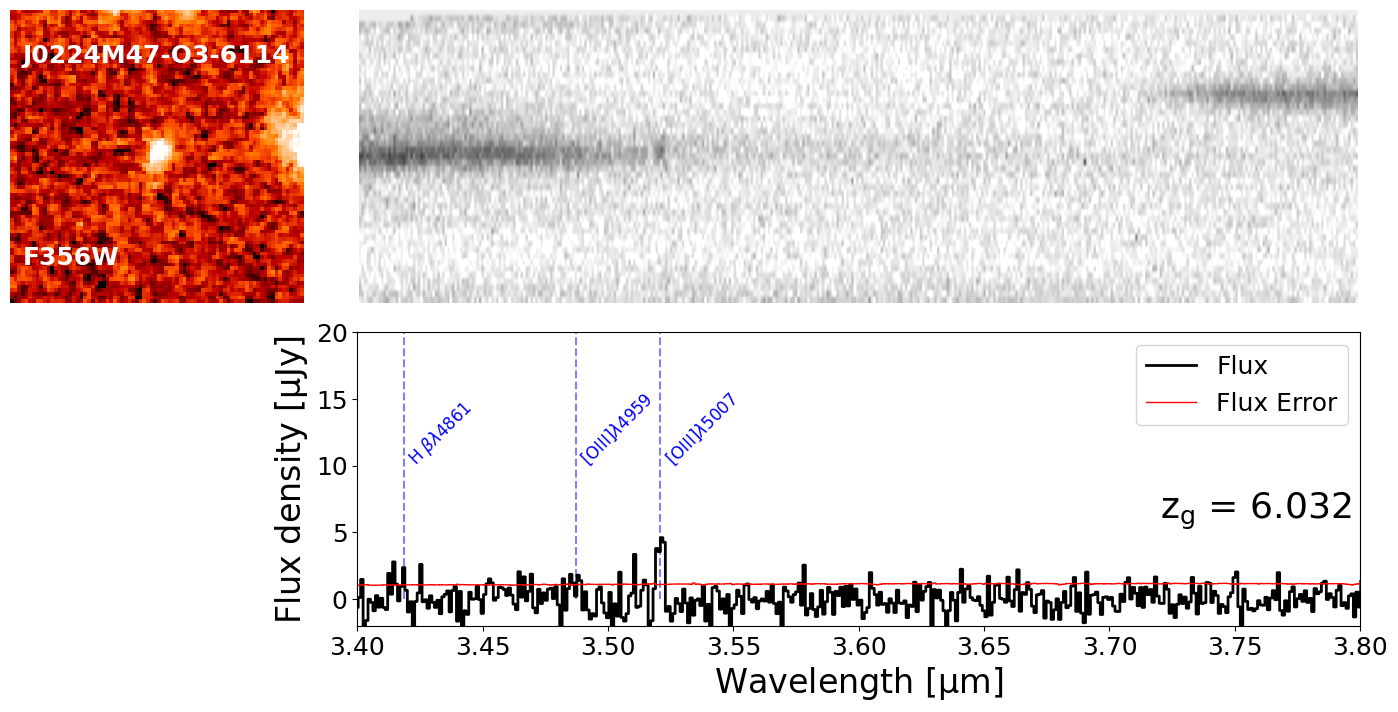}{0.5\textwidth}{}
\fig{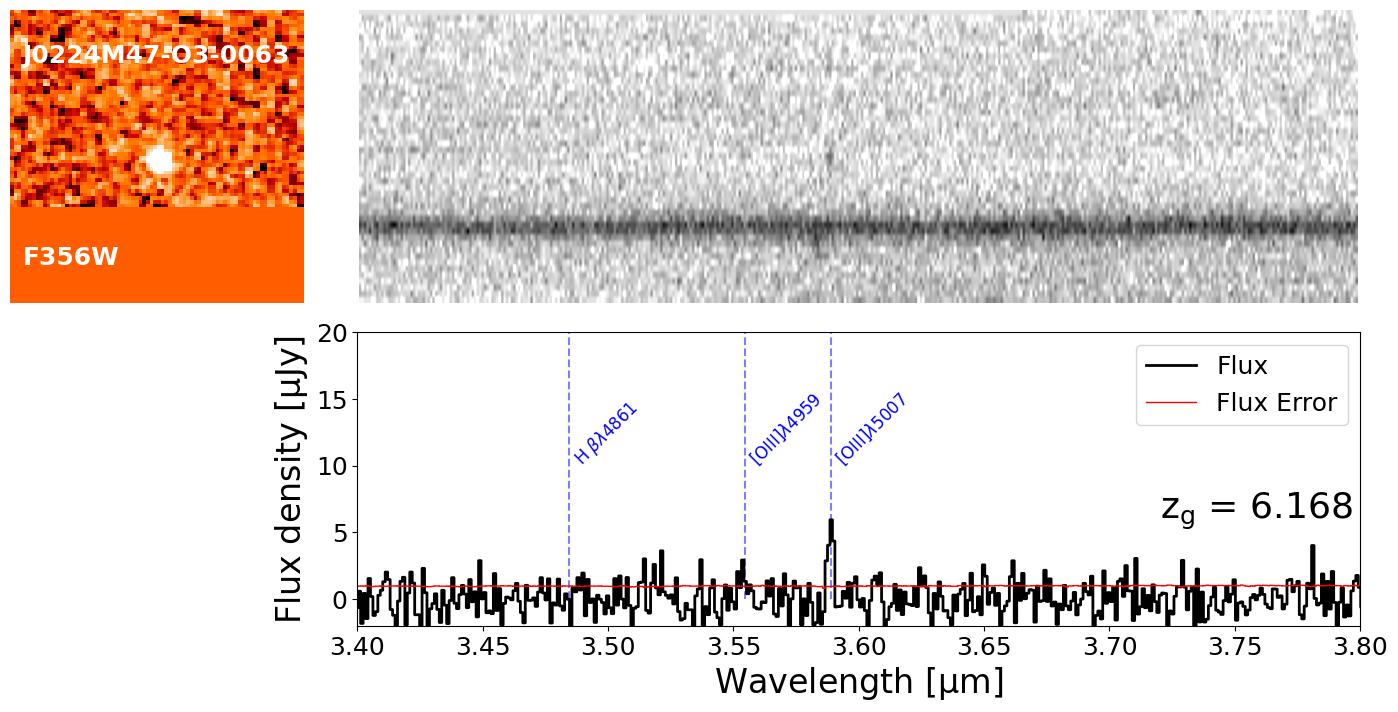}{0.5\textwidth}{}
}

\gridline{
\fig{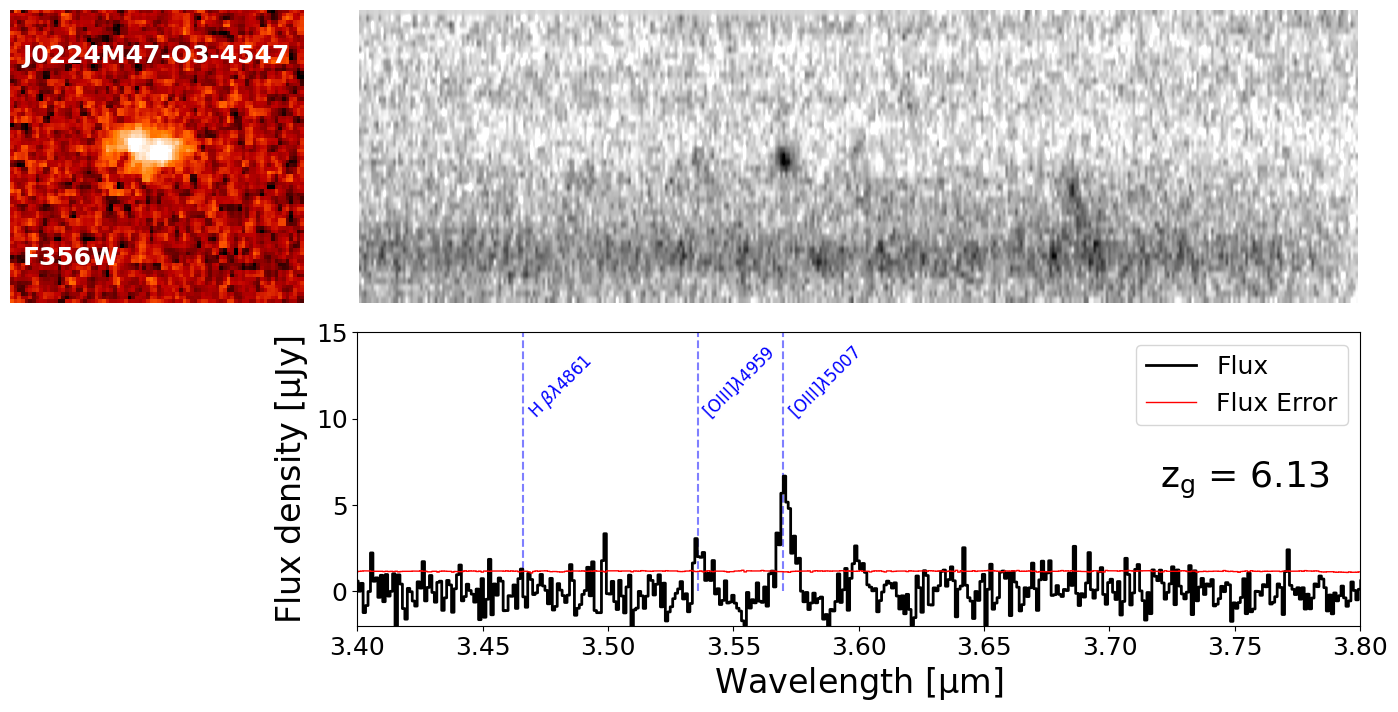}{0.5\textwidth}{}
\fig{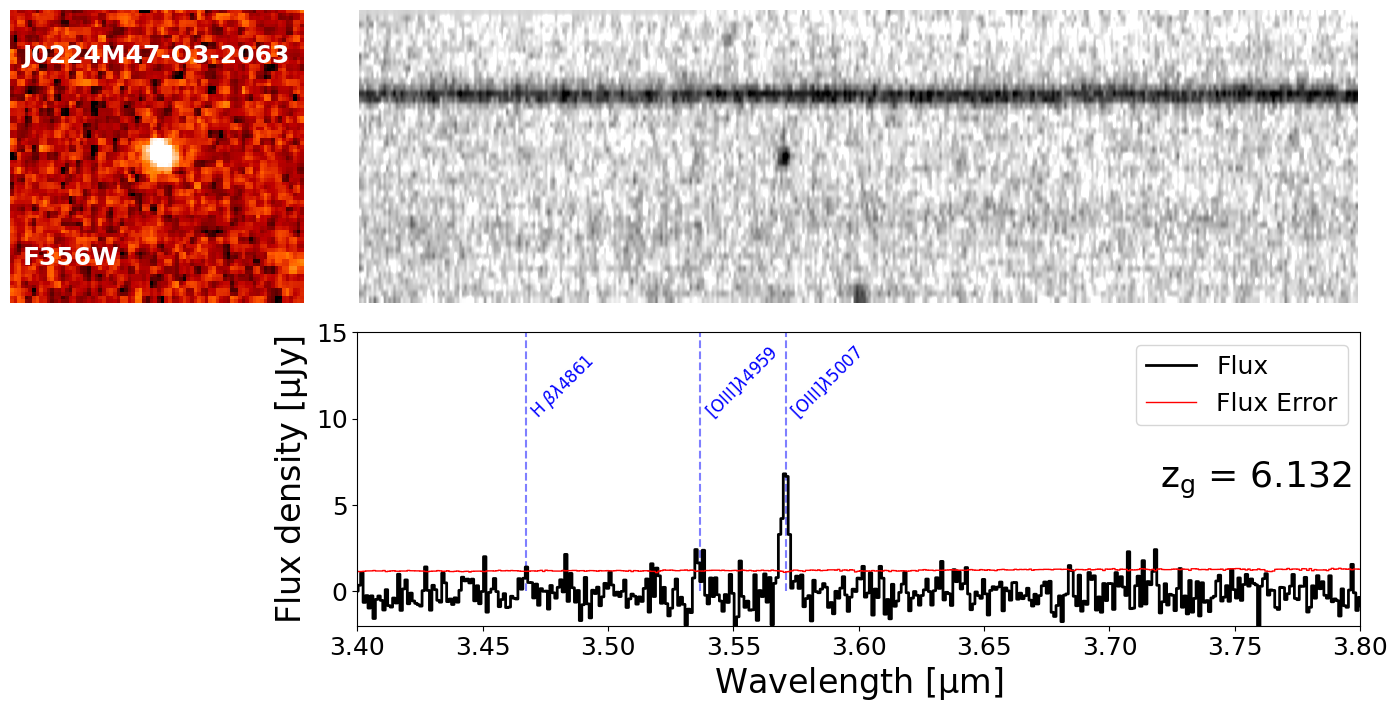}{0.5\textwidth}{}
}

\gridline{
\fig{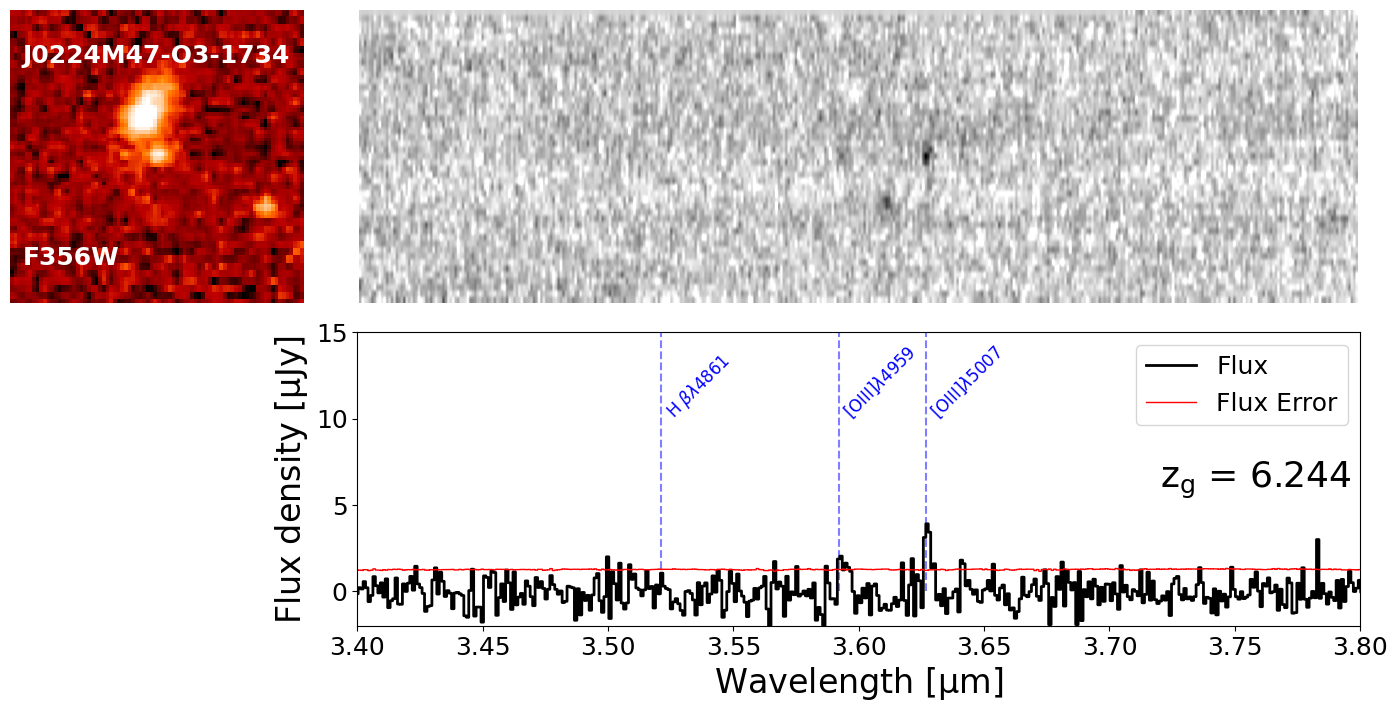}{0.5\textwidth}{}
\fig{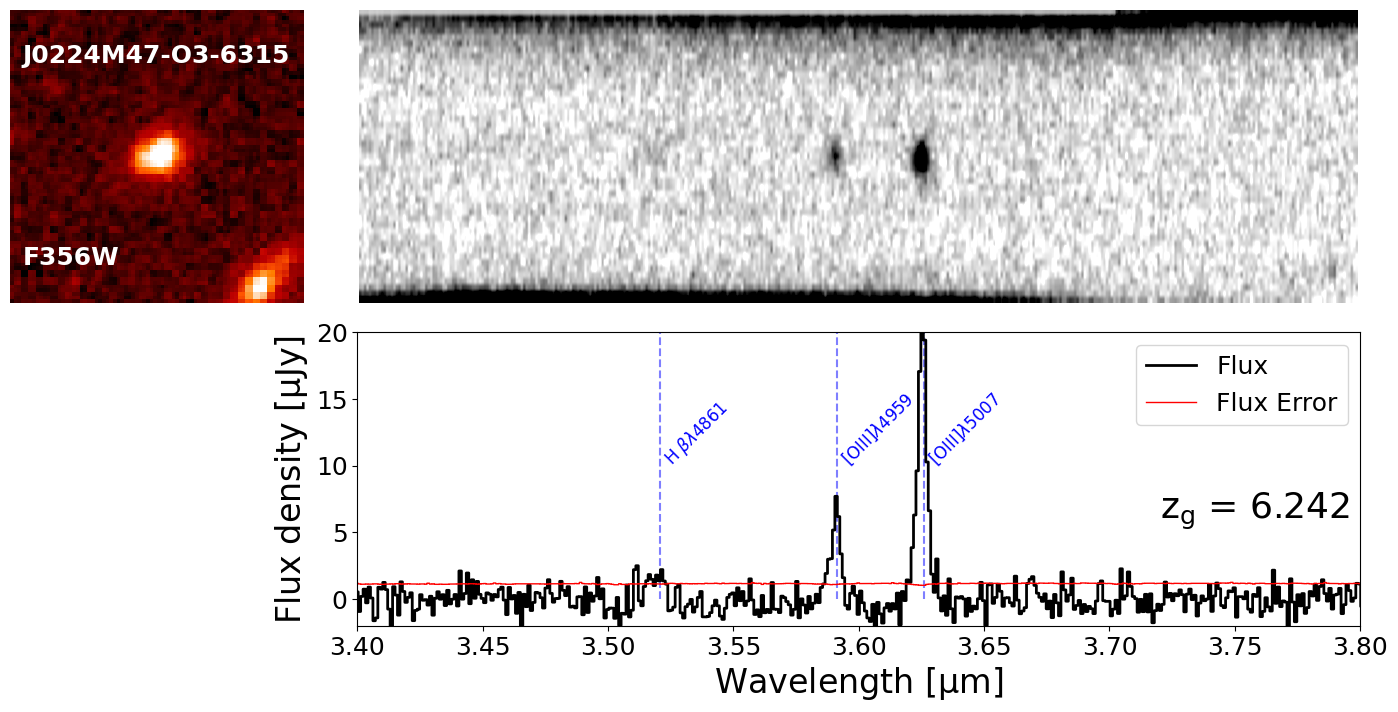}{0.5\textwidth}{}
}

\caption{All the absorbing related \oiii-emitting galaxy candidates.}\label{fig:o3_spec}
\end{figure*}


%
%
%

\begin{figure*}[htb!]
\gridline{
\fig{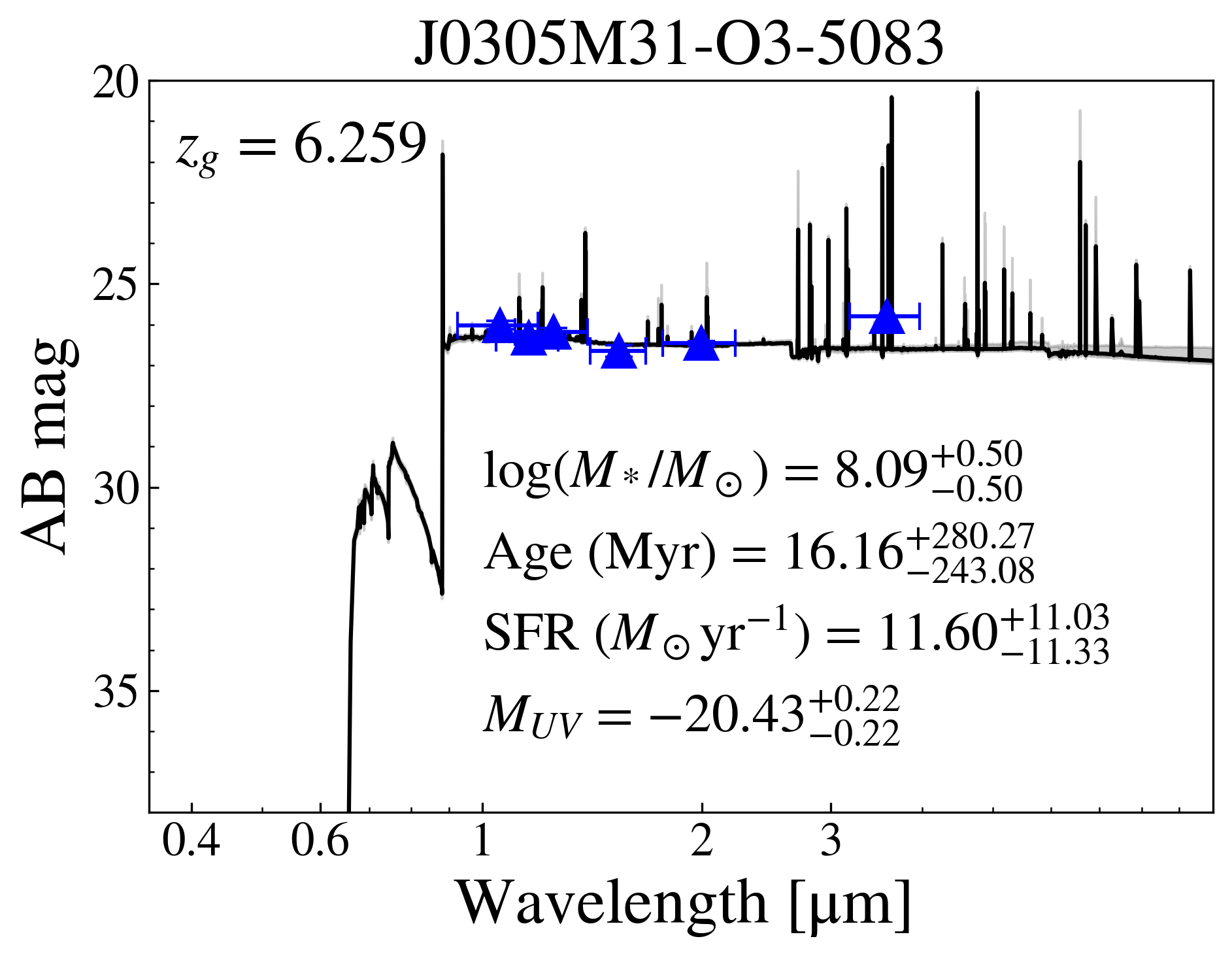}{0.3\textwidth}{}
\fig{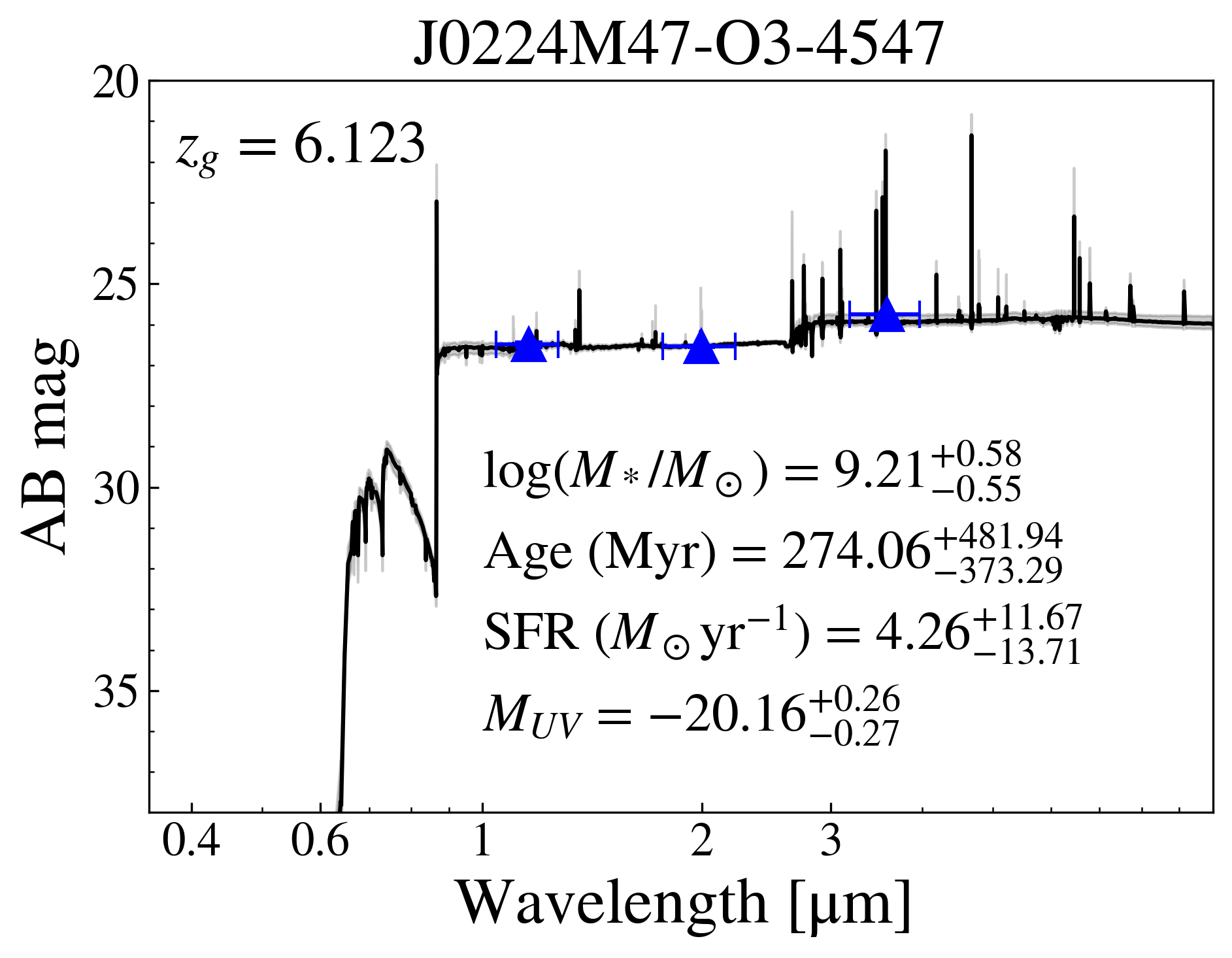}{0.3\textwidth}{}
\fig{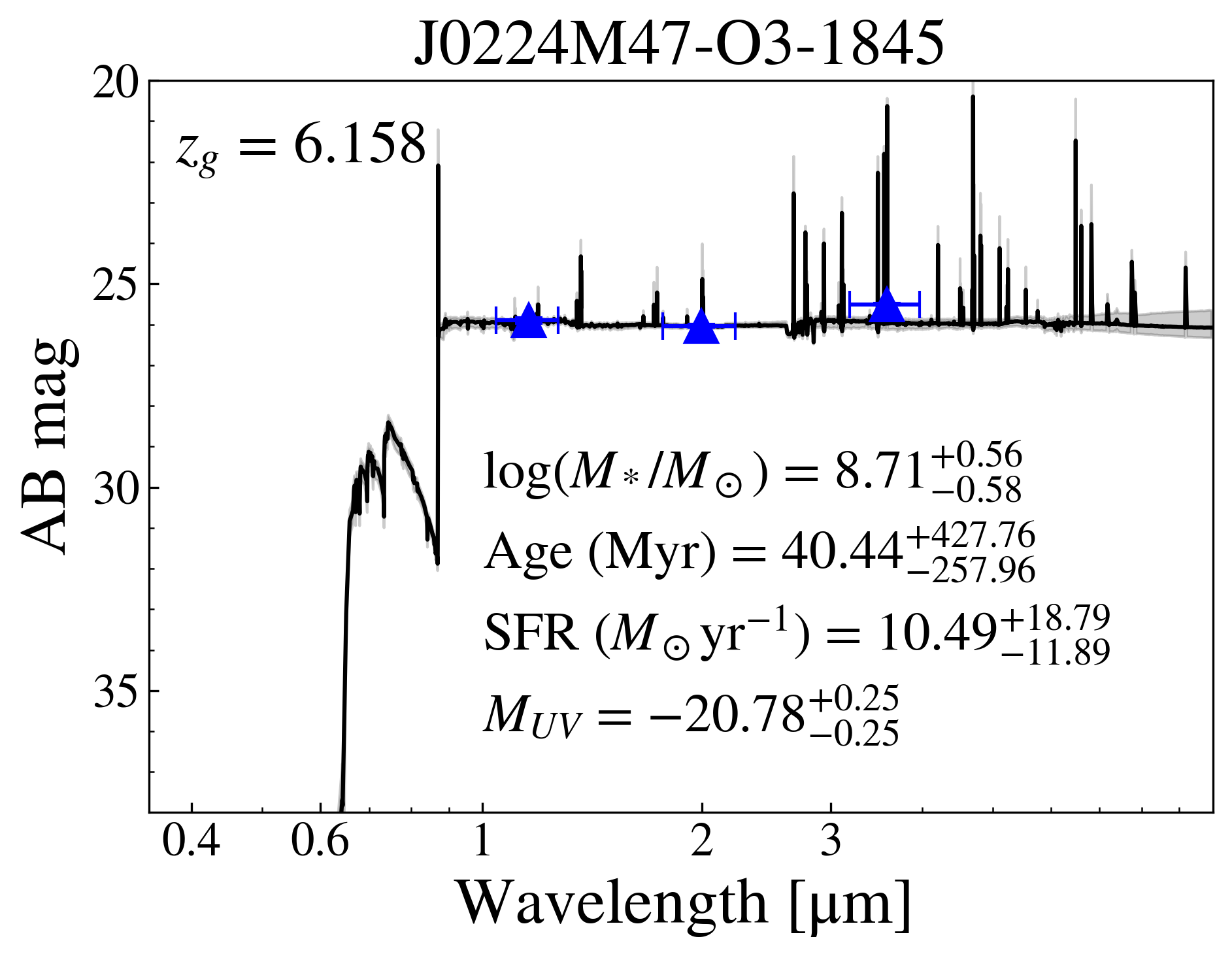}{0.3\textwidth}{}
}
\gridline{
\fig{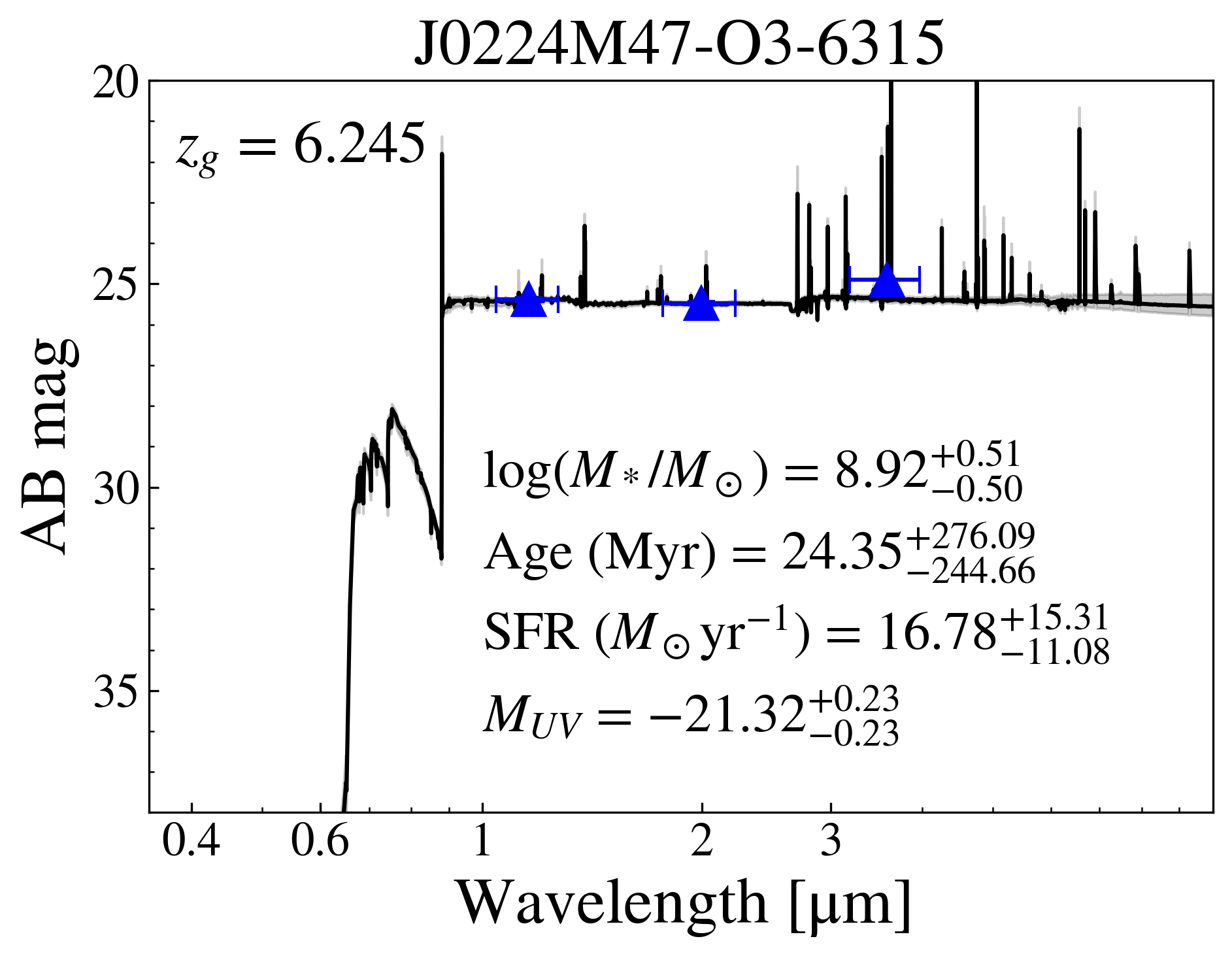}{0.3\textwidth}{}
\fig{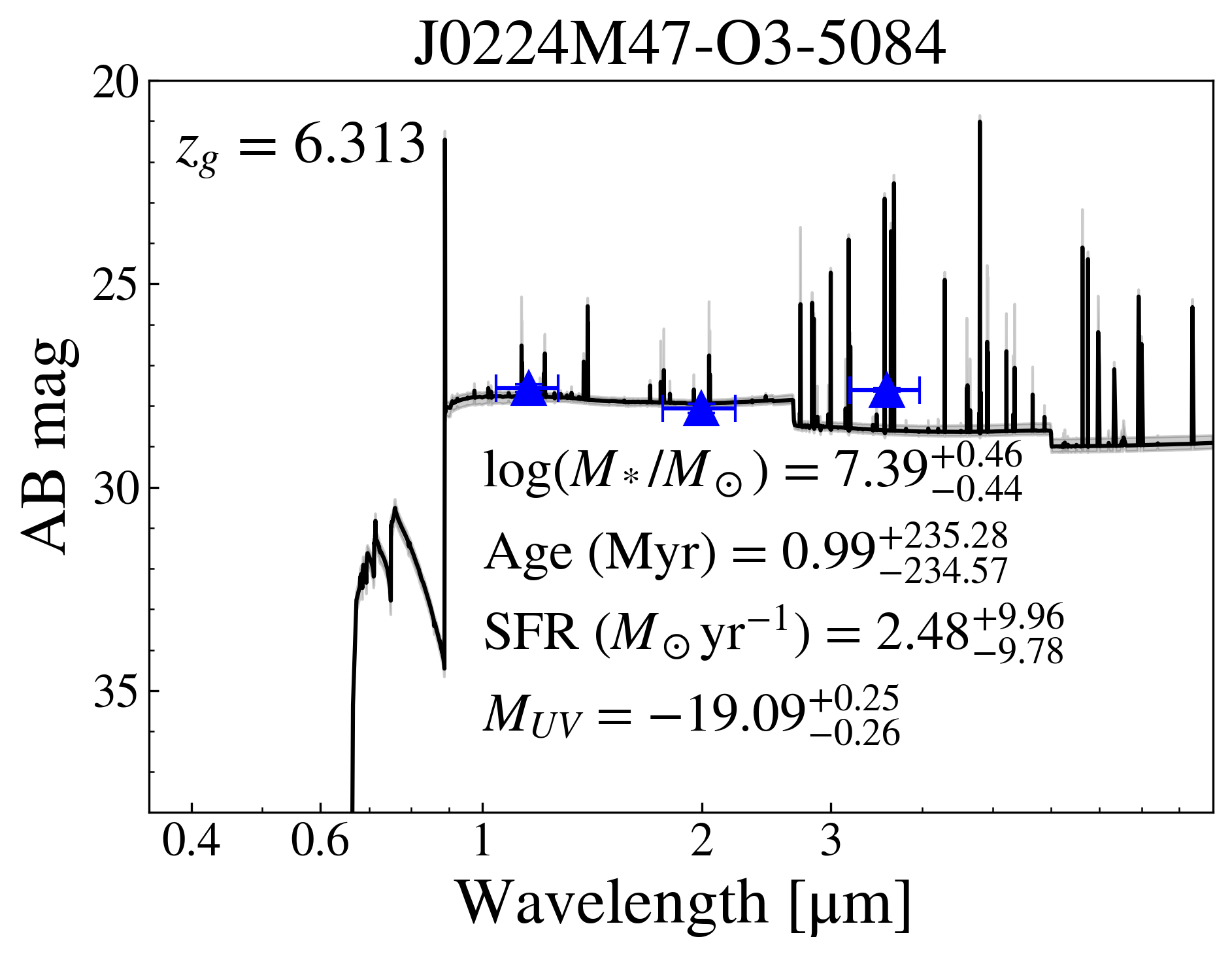}{0.3\textwidth}{}
\fig{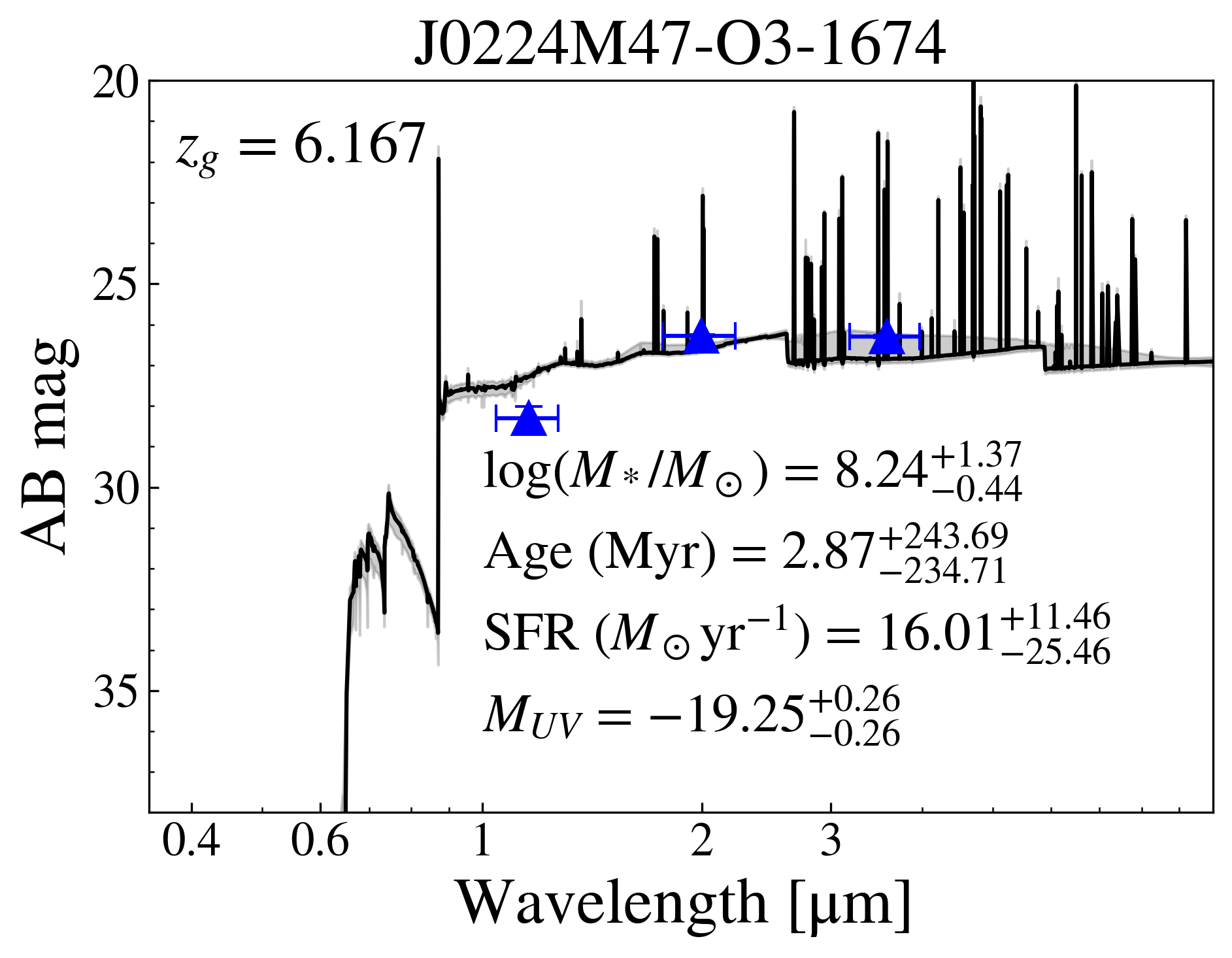}{0.3\textwidth}{}
}

\gridline{
\fig{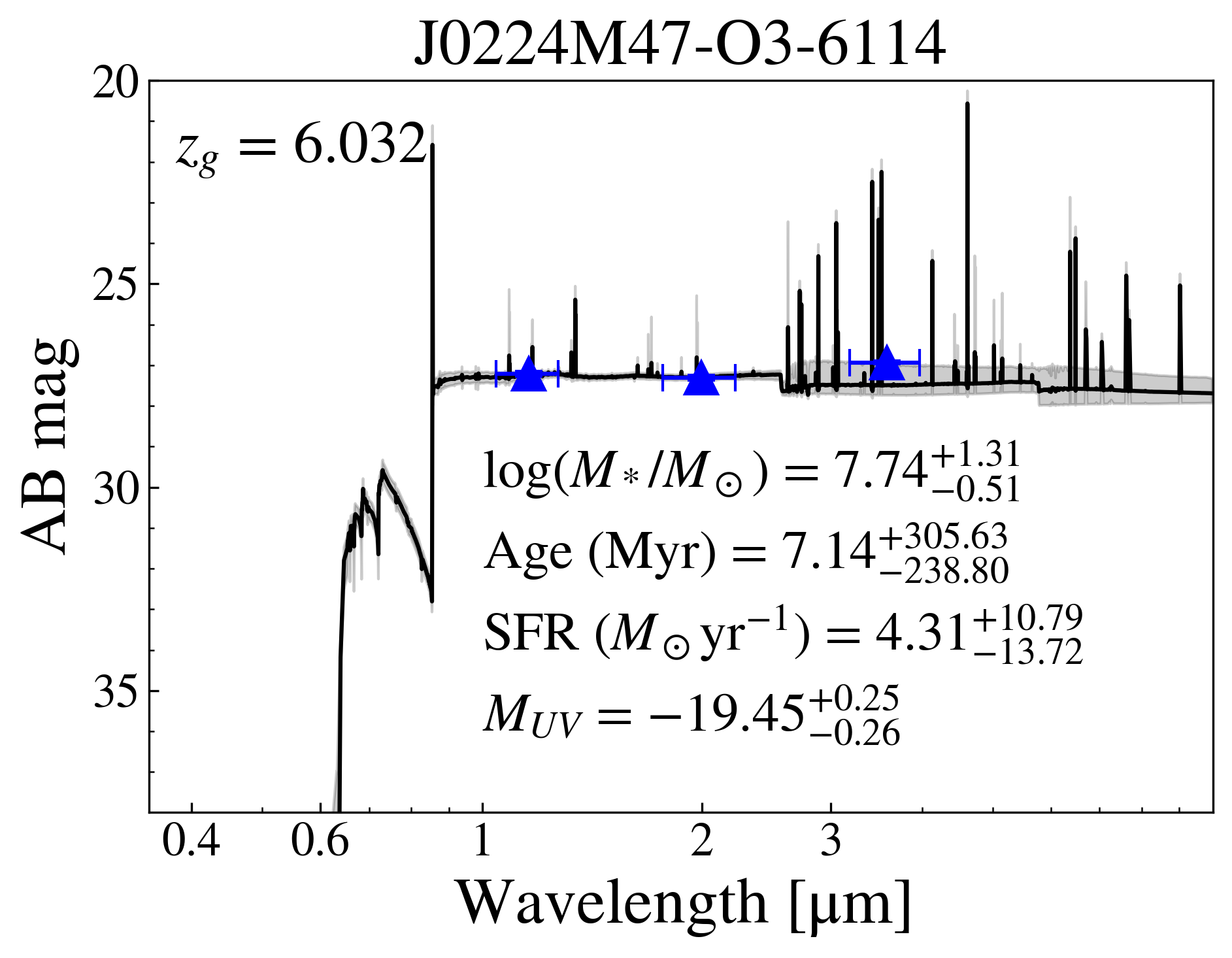}{0.3\textwidth}{}
\fig{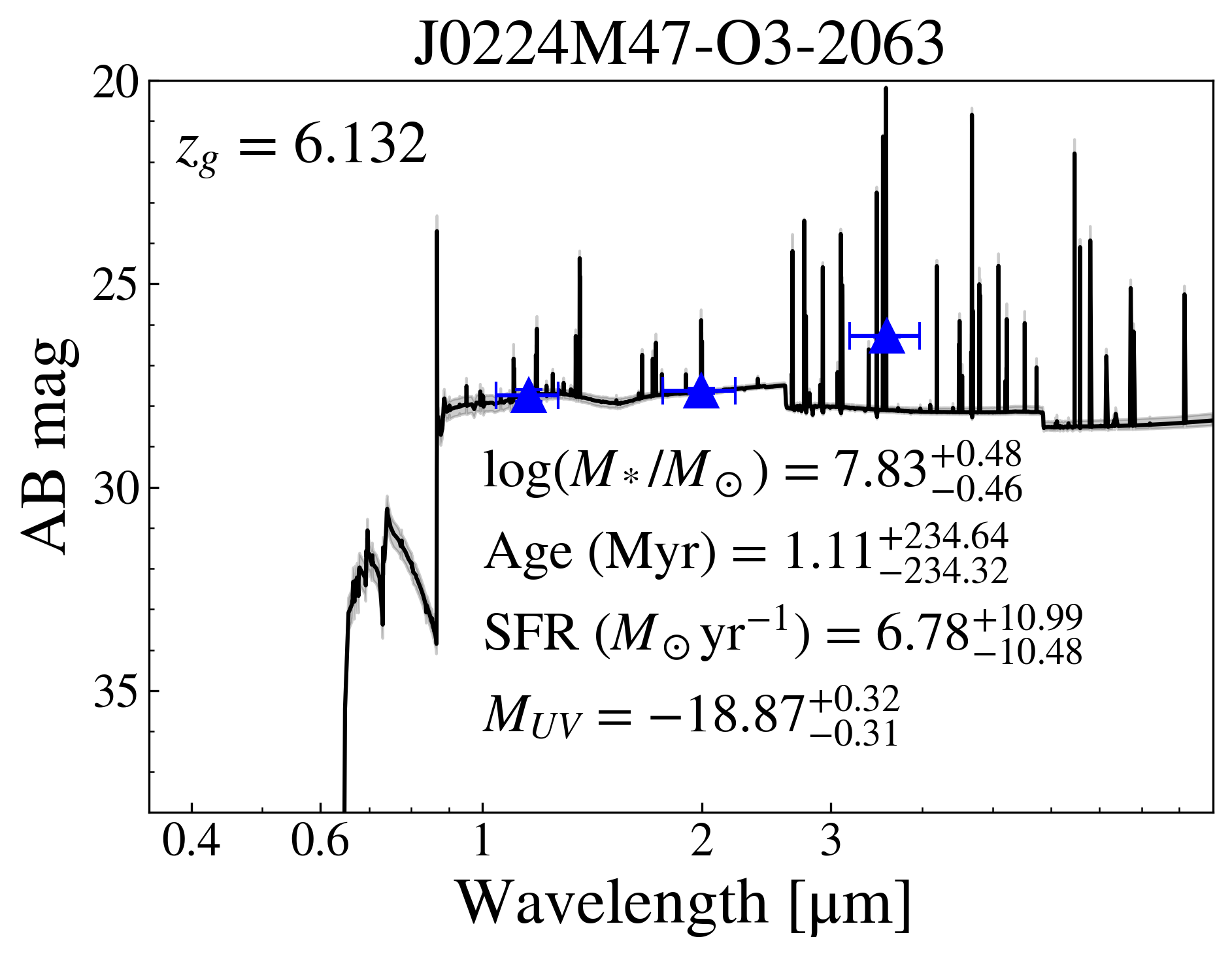}{0.3\textwidth}{}
\fig{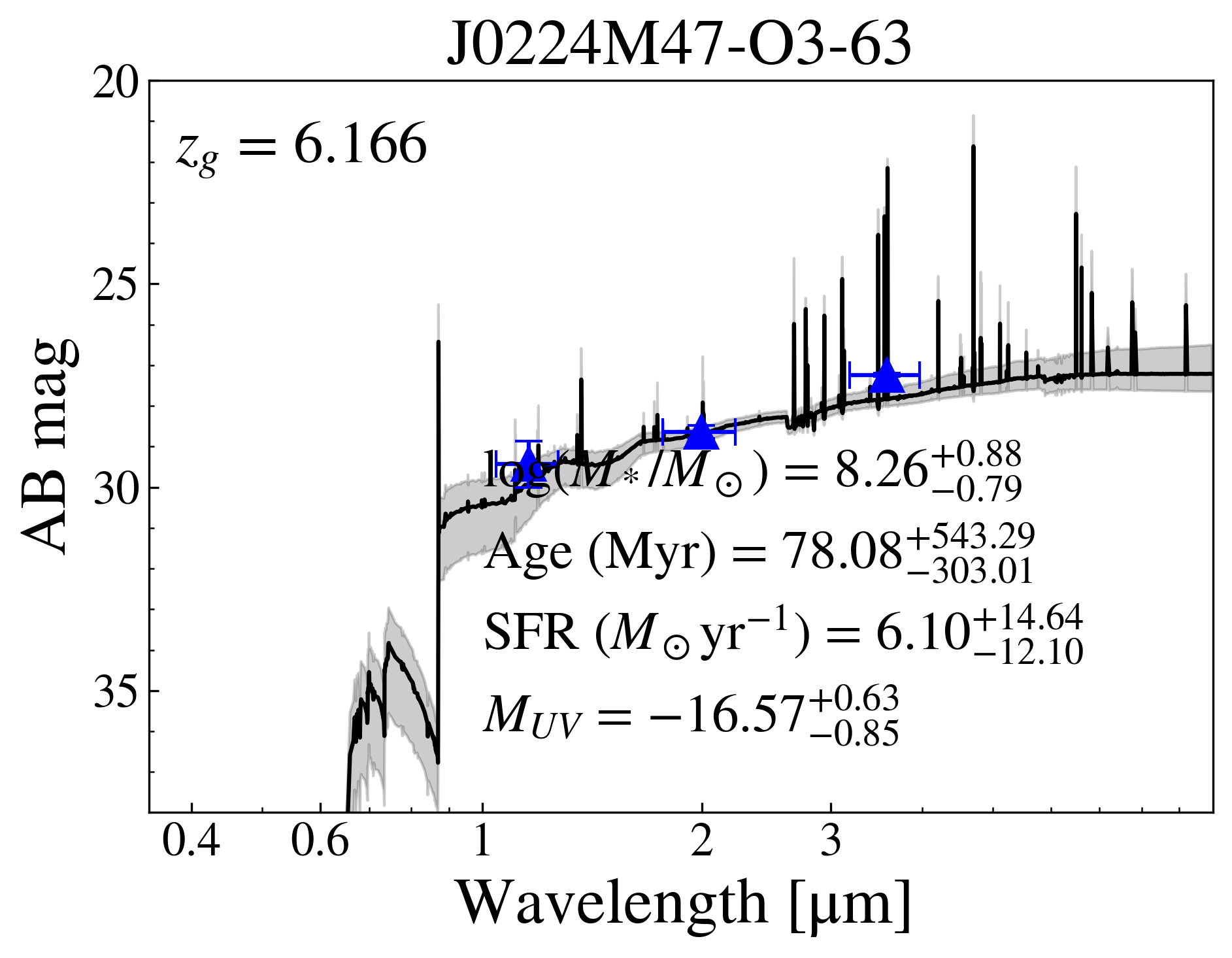}{0.3\textwidth}{}
}

\gridline{
\fig{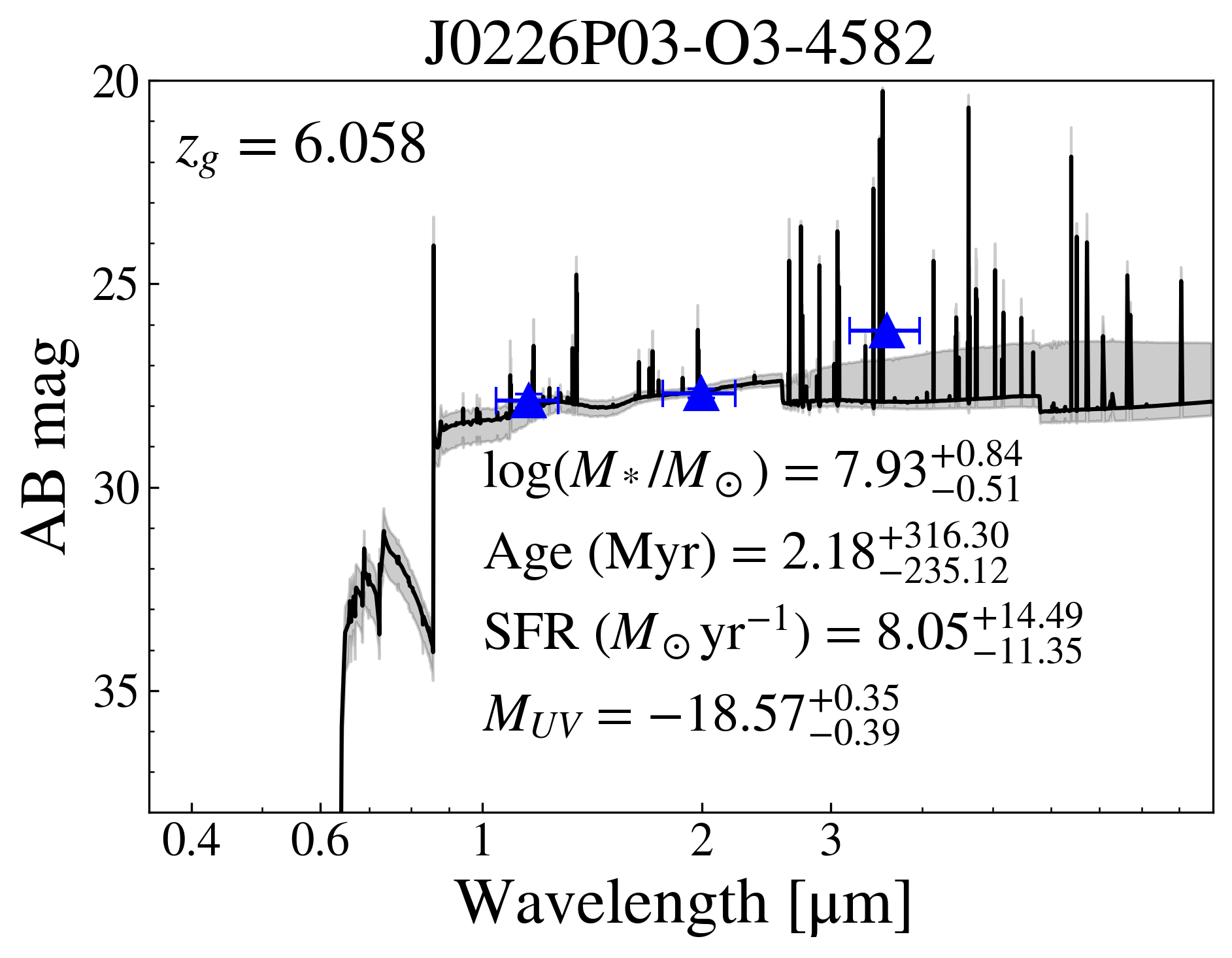}{0.3\textwidth}{}
\fig{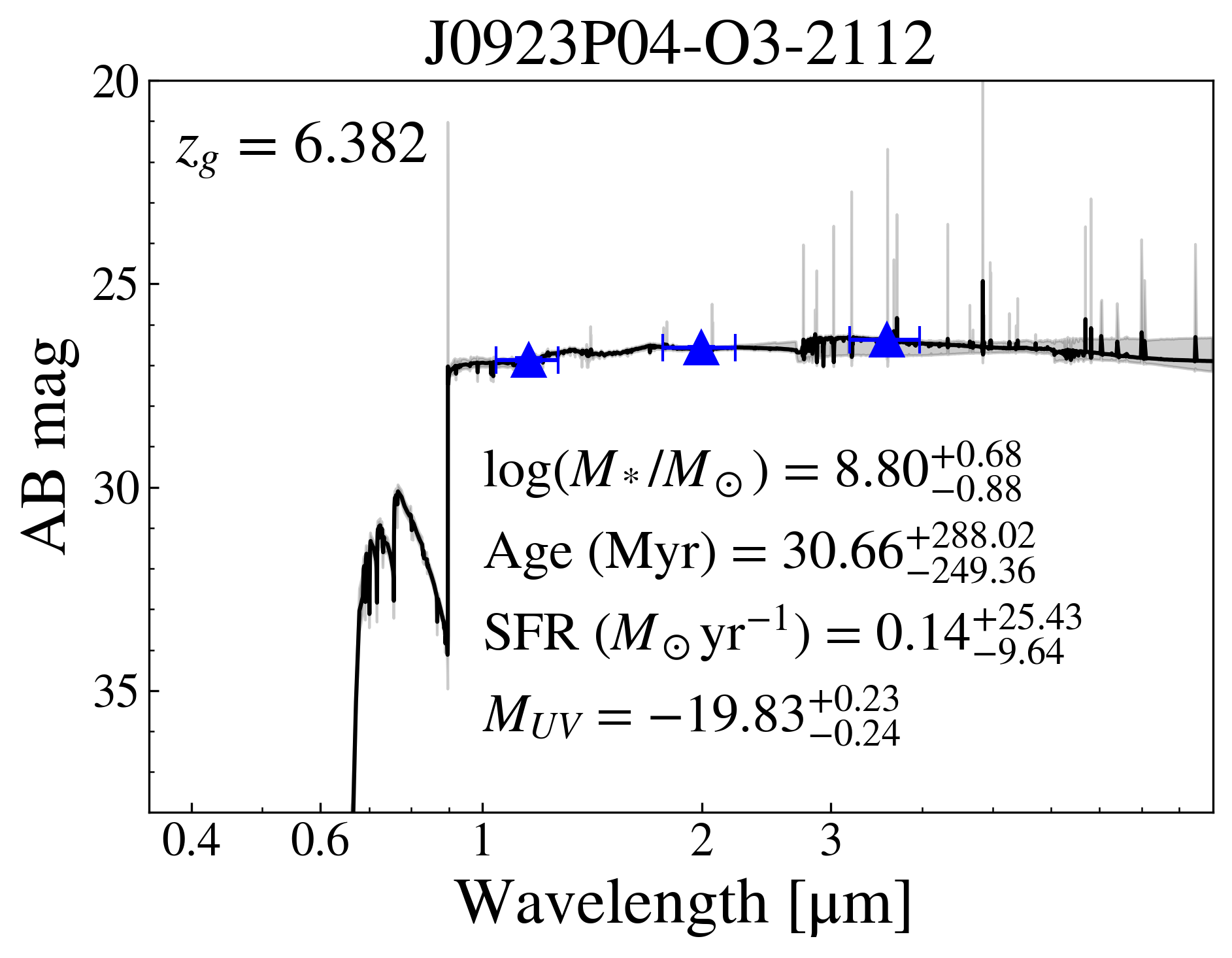}{0.3\textwidth}{}
\fig{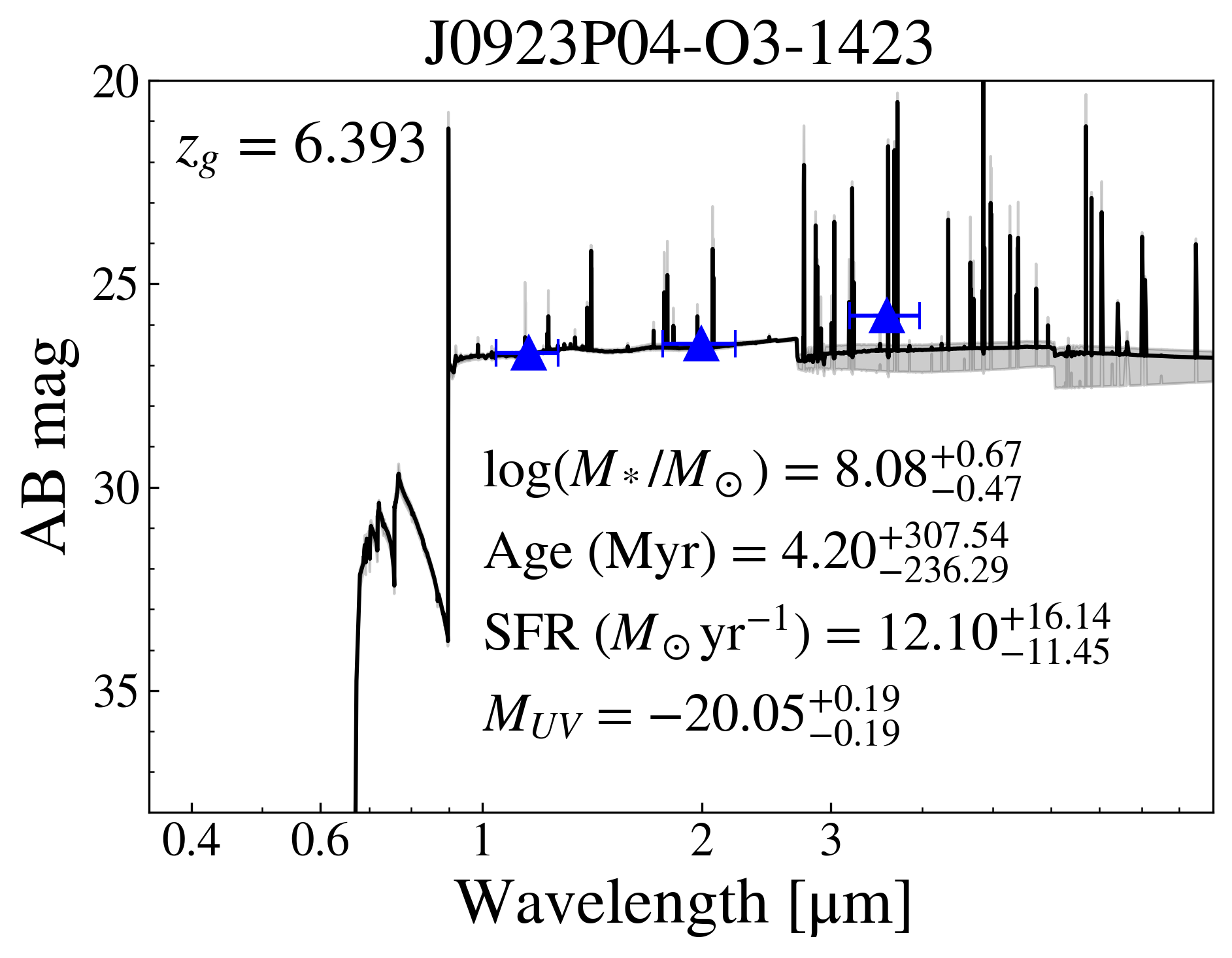}{0.3\textwidth}{}
}

\caption{SED fitting of all detected absorbing gas associated galaxy candidates. The uncertainty is derived from the BEAGLE fitting results, taking into account the uncertainty when compared with the JAGUAR mock catalog \citep{jaguar}. }\label{fig:sed}
\end{figure*}

\end{document}